\pdfoutput=1 
\documentclass[dvipsnames]{elsarticle}
\usepackage{hyperref}
\usepackage{graphicx}  
\usepackage{dcolumn}   
\usepackage{bm}        
\usepackage{amssymb}   
\usepackage{amsfonts}
\usepackage{graphics}
\usepackage{grffile}   
\usepackage{epsfig}
\usepackage[usenames]{color}
\usepackage[normalem]{ulem} 
\usepackage{color}
\usepackage[utf8]{inputenc}
\usepackage[T1]{fontenc}
\usepackage{amsmath}
\usepackage{booktabs}
\usepackage{svg}
\usepackage{dsfont}
\usepackage{enumitem}

\usepackage{xcolor}
\usepackage{tikz}
\usepackage{environ}
\makeatletter
    \newsavebox{\measure@tikzpicture}
        \NewEnviron{scaletikzpicturetowidth}[1]{%
            \def\tikz@width{#1}%
            \def\tikzscale{1}\begin{lrbox}{\measure@tikzpicture}%
            \BODY
            \end{lrbox}%
            \pgfmathparse{#1/\wd\measure@tikzpicture}%
            \edef\tikzscale{\pgfmathresult}%
            \BODY
        }
\makeatother
\usetikzlibrary{math}
\usetikzlibrary{calc}
\usetikzlibrary{decorations.pathreplacing, decorations.markings}
\usetikzlibrary{patterns, arrows, arrows.meta}

\usepackage{subcaption}
\usepackage{siunitx}

\DeclareSIUnit{\ele}{\mbox{$e^{\text{-}}$}}
\DeclareSIUnit[number-unit-product = ]{\percent}{\%}
\DeclareSIUnit{\raddose}{rad}
\DeclareSIUnit{\tid}{\mbox{\si{\kilo\raddose}}}
\DeclareSIUnit{\niel}{\mbox{\SI{1}{\mega\electronvolt}\ n$_{\mathrm{eq}}$\ \si{\per\cm\squared}}}
\DeclareSIUnit{\pixel}{pixel}

\newcommand{\Vsub}{V_{sub}}
\newcommand{\Vpwell}{V_{pwell}}

\newcommand{\Vb}{V_{casb}}
\newcommand{\Vn}{V_{casn}}
\newcommand{\Ir}{I_{reset}}
\newcommand{\Id}{I_{db}}
\newcommand{\Ib}{I_{bias}}
\newcommand{\Ibn}{I_{biasn}}

\journal{Nucl. Instrum. Methods Phys. Res. A}
\begin{document}%
\begin{frontmatter}


\title{Further Characterisation of Digital Pixel Test Structures Implemented in a 65~nm CMOS Process}


\author[1]{Gianluca Aglieri Rinella}
\author[20]{Nicole Apadula}
\author[13]{Anton Andronic}
\author[6]{Matias Antonelli}
\author[9,10]{Mauro Aresti}
\author[6]{Roberto Baccomi}
\author[2]{Pascal Becht}
\author[3,4]{Stefania Beole}
\author[21]{Marcello Borri}
\author[1,18]{Justus Braach}
\author[21]{Matthew Daniel Buckland\corref{cor1}}
\author[1]{Eric Buschmann}
\author[5,6]{Paolo Camerini}
\author[1]{Francesca Carnesecchi}
\author[1]{Leonardo Cecconi}
\author[17]{Edoardo Charbon}
\author[5,6]{Giacomo Contin}
\author[1]{Dominik Dannheim}
\author[1]{Joao de Melo}
\author[1,15]{Wenjing Deng}
\author[1]{Antonello di Mauro}
\author[1]{Jan Hasenbichler}
\author[1]{Hartmut Hillemanns}
\author[1,16]{Geun Hee Hong}
\author[7]{Artem Isakov}
\author[22]{Hangil Jang}
\author[1]{Antoine Junique}
\author[23]{Minjung Kim}
\author[1]{Alex Kluge}
\author[7]{Artem Kotliarov}
\author[7]{Filip K\v{r}\'i\v{z}ek}
\author[1,8]{Lukas Lautner}
\author[22]{Sanghoon Lim}
\author[1]{Magnus Mager}
\author[9,10]{Davide Marras}
\author[1]{Paolo Martinengo}
\author[2]{Silvia Masciocchi}
\author[2]{Marius Wilm Menzel}
\author[1]{Magdalena Munker}
\author[1,17]{Francesco Piro}
\author[6]{Alexandre Rachevski}
\author[1]{Karoliina Rebane}
\author[1]{Felix Reidt}
\author[11]{Roberto Russo}
\author[1]{Isabella Sanna}
\author[9,10]{Valerio Sarritzu}
\author[12]{Serhiy Senyukov}
\author[1]{Walter Snoeys}
\author[11]{Jory Sonneveld}
\author[1]{Miljenko \v{S}ulji\'c}
\author[1]{Peter Svihra}
\author[1,13]{Nicolas Tiltmann}
\author[5,24]{Vittorio Di Trapani}
\author[9,10]{Gianluca Usai}
\author[1]{Jacob Bastiaan Van Beelen}
\author[14,19]{Mirella Dimitrova Vassilev}
\author[14,19]{Caterina Vernieri}
\author[5,6]{Anna Villani}


\affiliation[1]{organization={European Organisation for Nuclear Research (CERN)}, city={Geneva}, country={Switzerland}}
\affiliation[20]{organization={Lawrence Berkeley National Laboratory}, city={Berkeley}, country={USA}}
\affiliation[23]{organization={University of California}, city={Berkeley}, country={USA}}
\affiliation[13]{organization={Universit\"at M\"unster}, city={M\"unster}, country={Germany}}
\affiliation[2]{organization={Ruprecht Karls Universit\"at Heidelberg}, city={Heidelberg}, country={Germany}}
\affiliation[3]{organization={Universit\`a degli studi di Torino}, city={Torino}, country={Italy}}
\affiliation[4]{organization={Istituto Nazionale di Fisica Nucleare (INFN), Sezione di Torino}, city={Torino}, country={Italy}}
\affiliation[18]{organization={Universit\"at Hamburg}, city={Hamburg}, country={Germany}}
\affiliation[5]{organization={Universit\`a degli studi di Trieste}, city={Trieste}, country={Italy}}
\affiliation[6]{organization={Istituto Nazionale di Fisica Nucleare (INFN), Sezione di Trieste}, city={Trieste}, country={Italy}}
\affiliation[24]{organization={Elettra Sincrotrone S.C.p.A.}, city={Basovizza}, country={Italy}}
\affiliation[21]{organization={STFC Daresbury Laboratory}, city={Daresbury}, country={United Kingdom}}
\affiliation[17]{organization={EPFL}, city={Lausanne}, country={Switzerland}}
\affiliation[15]{organization={Central China Normal University (CCNU)}, city={Wuhan}, country={China}}
\affiliation[16]{organization={Yonsei University}, city={Seoul}, country={Republic of Korea}}
\affiliation[22]{organization={Pusan National University}, city={Pusan}, country={Republic of Korea}}
\affiliation[7]{organization={Czech Academy of Sciences}, city={Prague}, country={Czechia}}
\affiliation[8]{organization={Technische Universit\"at M\"unchen}, city={M\"unchen}, country={Germany}}
\affiliation[9]{organization={Universit\`a degli studi di Cagliari}, city={Cagliari}, country={Italy}}
\affiliation[10]{organization={Istituto Nazionale di Fisica Nucleare (INFN), Sezione di Cagliari}, city={Cagliari}, country={Italy}}
\affiliation[11]{organization={Nikhef National institute for subatomic physics}, city={Amsterdam}, country={Netherlands}}
\affiliation[12]{organization={Centre National de la Recherche Scientifique}, city={Strasbourg}, country={France}}
\affiliation[14]{organization={SLAC}, city={Menlo Park}, country={California, USA}}
\affiliation[19]{organization={Stanford University}, city={Stanford}, country={California, USA}}

\cortext[cor1]{Corresponding author}


\pagebreak

\begin{abstract}

The next generation of MAPS for future tracking detectors will have to meet stringent requirements placed on them. 
One such detector is the ALICE ITS3 that aims to be very light at \qty{0.07}{\percent}~X/X$_{0}$ per layer and have a low power consumption in the active area of \qty{40}{\milli\watt\per\cm^2} by implementing wafer-scale MAPS bent into cylindrical half layers.
To address these challenging requirements, the ALICE ITS3 project, in conjunction with the CERN EP R\&D on monolithic pixel sensors, proposed the Tower Partners Semiconductor Co.\ \SI{65}{\nm} CMOS process as the starting point for the sensor. After the initial results confirmed the detection efficiency and radiation hardness, the choice of the technology was solidified by demonstrating the feasibility of operating MAPS in low-power consumption regimes, $<\qty{50}{\milli\watt\per\cm^2}$, while maintaining high-quality performance. This was shown through a detailed characterisation of the Digital Pixel Test Structure (DPTS) prototype exposed to X-rays and ionising beams, and the results are presented in this article. Additionally, the sensor was further investigated through studies of the fake-hit rate, the linearity of the front-end in the range \SIrange{1.7}{28}{keV}, the performance after ionising irradiation, and the detection efficiency of inclined tracks in the range \SIrange{0}{45}{\degree}.


\end{abstract}



\end{frontmatter}

%
%

\tikzset{
	beamarrow/.style={
		decoration={
			markings,mark=at position 1 with 
			{\arrow[scale=2,>=stealth]{>}}
		},postaction={decorate}
	}
}
\tikzset{
	pics/.cd,
	vector out/.style={
		code={
		\draw[#1, thick] (0,0)  circle (0.15) (45:0.15) -- (225:0.15) (135:0.15) -- (315:0.15);
		}
	}
}
\tikzset{
	pics/.cd,
	vector in/.style={
		code={
		\draw[#1, thick] (0,0)  circle (0.15);
		 \fill[#1] (0,0)  circle (.05);
		 }
	}
}
\tikzset{
	global scale/.style={
		scale=#1,
		every node/.style={scale=#1}
	}
}
\def\centerarc[#1] (#2)(#3:#4:#5) 
	 { \draw[#1] ($(#2)+({#5*cos(#3)},{#5*sin(#3)})$) arc (#3:#4:#5); }

\section{Introduction}
\label{sec:intro}

Monolithic Active Pixel Sensors (MAPS) incorporate the charge-sensitive volume and the read-out circuitry in one piece of silicon and can be fabricated in standard CMOS processes. This allows for a sensor that has promising characteristics for tracking detectors, such as low capacitance, low mass and low power consumption. MAPS were intensively developed and used over the last decade \cite{SNOEYS2023168678}. MAPS have demonstrated their suitability for operation in high-energy physics experiment vertex and tracking detectors~\cite{star}. The most recent implementation of a large-area MAPS detector was the Inner Tracking System 2 (ITS2) for the ALICE experiment at CERN~\cite{ls2paper,its2conf}. The ITS2 uses the ALPIDE chip~\cite{ALPIDE-proceedings-2,ALPIDE-proceedings-1}, which was fabricated in the TowerJazz \qty{180}{\nm} CMOS process and showed excellent performance in terms of detection efficiency (\qty{\gg 99}{\percent}) and spatial resolution (about \qty{5}{\um}).

The ITS2 will undergo an upgrade called the ITS3, where the three innermost tracking layers will be replaced during the LHC Long Shutdown 3 (2026--2028) \cite{BUCKLAND2022166875}. The ITS3 employs wafer-scale MAPSs with a length of \qty{260}{\mm} that are thinned to $\leq\qty{50}{\um}$ to reduce the material in the detector. These sensors are required to have power consumption below \qty{40}{\milli\watt\per\cm^2} in the active area so that operation at temperatures achievable by air cooling $O(\qty{+25}{\degreeCelsius})$ is possible~\cite{The:2890181}. Due to the challenging requirements imposed on the next generation of MAPS detectors, the Tower Partners Semiconductor Co.~(TPSCo) \SI{65}{\nm} CMOS imaging process~\cite{tower} was chosen as the starting point for sensor development. The first submission in the \qty{65}{\nm} CMOS process, in conjunction with the CERN~EP~R\&D on monolithic pixel sensors~\cite{eprnd}, was called Multi-Layer Reticle 1 (MLR1). It contained many different test structures to fully explore the feasibility of using the CMOS process for use in future detectors~\cite{Buckland_2024}.

With the first set of results from the MLR1 demonstrating the detection efficiency and radiation hardness of the \qty{65}{\nm} technology~\cite{AGLIERIRINELLA2023168589,AGLIERIRINELLA2024169896}, further investigations into the performance of the technology were carried out. Therefore, a MAPS prototype called the Digital Pixel Test Structure (DPTS) was characterised by fluorescence X-rays and ionising particles from a beam, the results of which are presented in this paper. 

Since one of the crucial requirements for the next generation of MAPS is reduced power usage, a particular focus was placed on the operation and performance of the sensor in low-power settings. Given that the active region of the final ITS3 stitched sensor will be a complex design with many components, and the DPTS is a simpler design, a comparison of the active region power consumption is not appropriate. Instead, because the two chip designs have similar pixel architectures, the pixel matrix power consumption will be used for a more representative comparison. The target for the pixel matrix power consumption of the ITS3 sensor is \qty{15}{{\milli\watt\per\cm^2}} for a pixel pitch of $20.8\times22.8$~\si{\um} \cite{The:2890181}. In addition, the performance after ionising irradiation, a study of the Fake-Hit Rate (FHR), the evaluation of the linearity of the front-end, and the response to inclined tracks were also examined.

\section{The DPTS chip}
\label{sec:dpts_chip}

The DPTS sensor measures \qtyproduct{1.5x1.5}{\mm}, features a \numproduct{32x32} pixel matrix with a pitch of \qty{15}{\um} and contains full-CMOS circuitry~\cite{AGLIERIRINELLA2023168589, 10196055}. This allows a digital front-end in each pixel, including an amplifier, shaper, and discriminator, as detailed in~\cite{AGLIERIRINELLA2023168589}. Furthermore, a test circuitry that injects an externally controlled test charge is also present in the pixel. The sensor can be configured by the shift register located on the periphery, allowing pixel masking and pixel selection for pulsing. The sensor is controlled by external reference currents and voltages and is read out via a Current-Mode Logic (CML) output~\cite{cecconi-twepp}. All pixels are read out simultaneously via a differential digital output that encodes the pixel position and Time-over-Threshold (ToT).

The steady-state front-end power consumption of a pixel is given by the currents $\Ib$, $\Ibn$, $\Id$, and $\Ir$ multiplied by the supply voltage (\qty{+1.2}{\V}). Since the reset current is orders of magnitude lower than the other currents (\si{\pA} compared to \si{nA}) and the $\Id$ current in the discriminator branch flows only while the discriminator is active, the total current of the pixel is given by the main branch, $\Ib$. Thus, the lowest possible power consumption is \qty{12}{\nano\watt} per pixel due to \qty{10}{\nA} being the lowest operational $\Ib$ setting.

\section{Laboratory measurements}
\label{sec:lab_meas}

\subsection{Data acquisition setup}
\label{sec:lab_daq}

The biases and control signals are supplied to the DPTS chip via a custom-made setup, allowing the control of the in-pixel front-end. The chip biases are set to the so-called ``nominal'' values defined by the design operating point, as described in Ref.~\cite{AGLIERIRINELLA2023168589}, unless otherwise stated or varied in the measurement. For measurements that varied the reverse bias ($\Vsub$ and $\Vpwell$), both chip bias voltages that control the amplifier ($\Vb$ and $\Vn$) were adjusted accordingly ~\cite{AGLIERIRINELLA2023168589}. The digital signals from the chip are read out via a CML output and recorded on an oscilloscope with a sampling rate of 5~GS/s and a bandwidth of \qty{500}{\MHz}.

\subsection{Performance at low power consumption}
\label{sec:lab_power}

As mentioned in Sec.~\ref{sec:dpts_chip}, the sole contributor to the DPTS pixel matrix power consumption is the chip current bias parameter called $\Ib$. Thus, the performance at low power consumption levels was investigated by measuring the pixel threshold spread (threshold RMS), pixel noise (width of the s-curve) and FHR (number of noise hits per pixel per second in the absence of external stimuli) at the $\Ib$ values of \qtylist[list-units=single]{10;20;30;50;100}{\nA}. These values correspond to a pixel matrix power consumption of \qtylist[list-units=single]{5.3;10.6;15.9;26.5;53}{\milli\watt\per\cm^2}, respectively. In contrast to the previous results~\cite{AGLIERIRINELLA2023168589}, where $\Ib$ was varied while keeping the other settings at their ``nominal'' values, in this study, the chip bias settings at each $\Ib$ were optimised to obtain the best performance. The optimised chip biases were $\Vn$, $\Ir$ and $\Id$. For comparison, the measurements at $\Ib=\qty{100}{\nA}$ are left at the ``nominal'' settings. 

\begin{figure}[!t]
    \centering
    \includegraphics[width=0.9\textwidth]
    {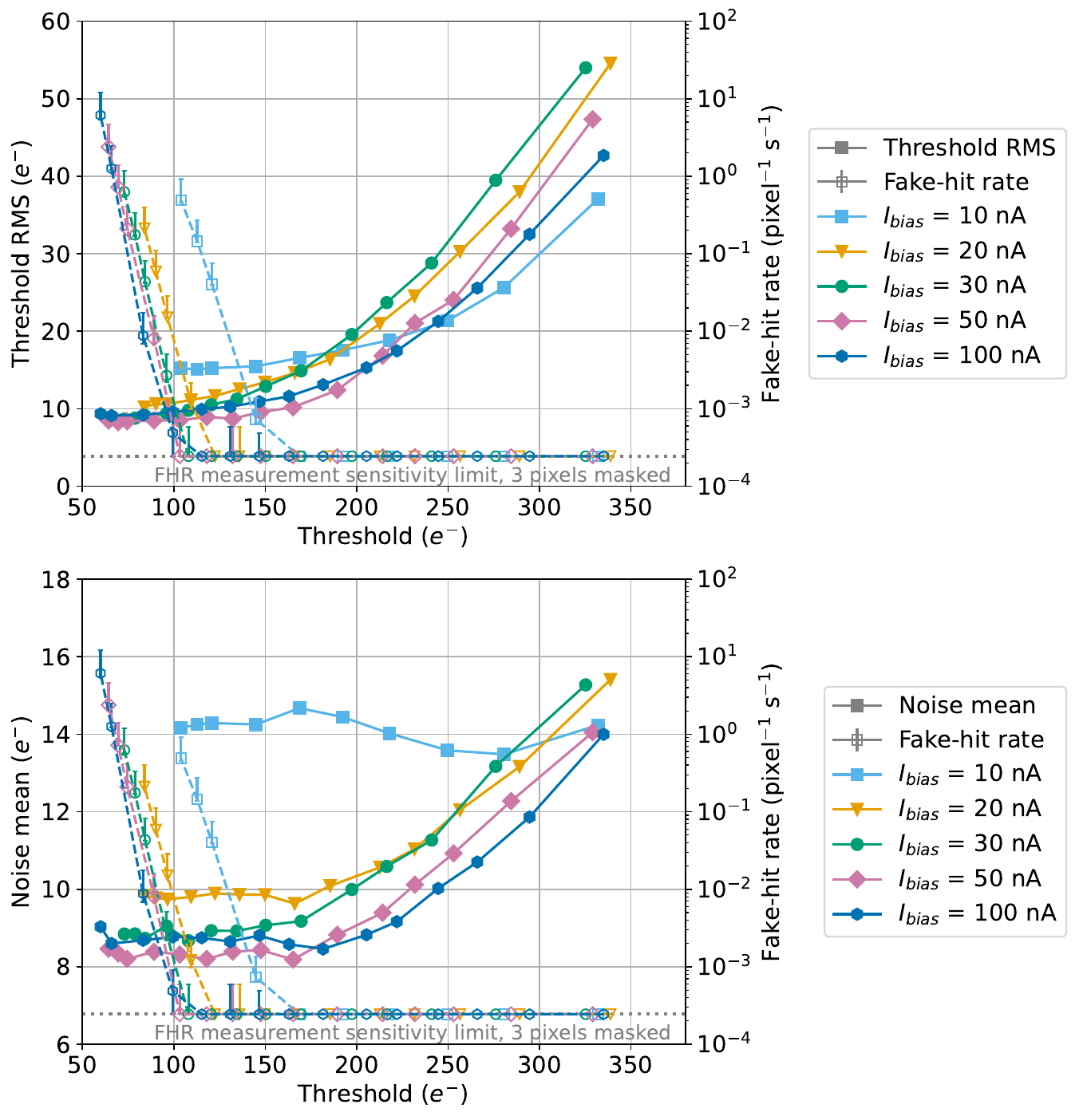}
    \caption{The threshold RMS and fake-hit rate vs. threshold (top) and the noise mean and fake-hit rate vs. threshold (bottom) at different $\Ib$ values. At each $\Ib$, the chip bias parameters were optimised, except $\Ib=\qty{100}{\nA}$, which is left at the ``nominal'' settings for comparison.}
    \label{fig:pwr_thr_noise_fhr}
\end{figure}

Figure~\ref{fig:pwr_thr_noise_fhr} shows the chip's performance as a function of the threshold for low $\Ib$ values. It can be seen that there is a small difference between the results at the thresholds $<\qty{200}{\ele}$ for $\Ib$ = \qtyrange{30}{100}{\nA}. This indicates some margin for lowering the power consumption with only a minimal impact on the performance. Whereas, for $\Ib=\qty{20}{\nA}$, the threshold RMS and the noise mean are slightly larger, and the onset of the FHR happens at slightly larger thresholds. This performance reduction continues for $\Ib=\qty{10}{\nA}$ but is more prominent, particularly for the noise mean and the FHR.

\begin{figure}[!t]
    \centering
    \includegraphics[width=0.99\textwidth]
    {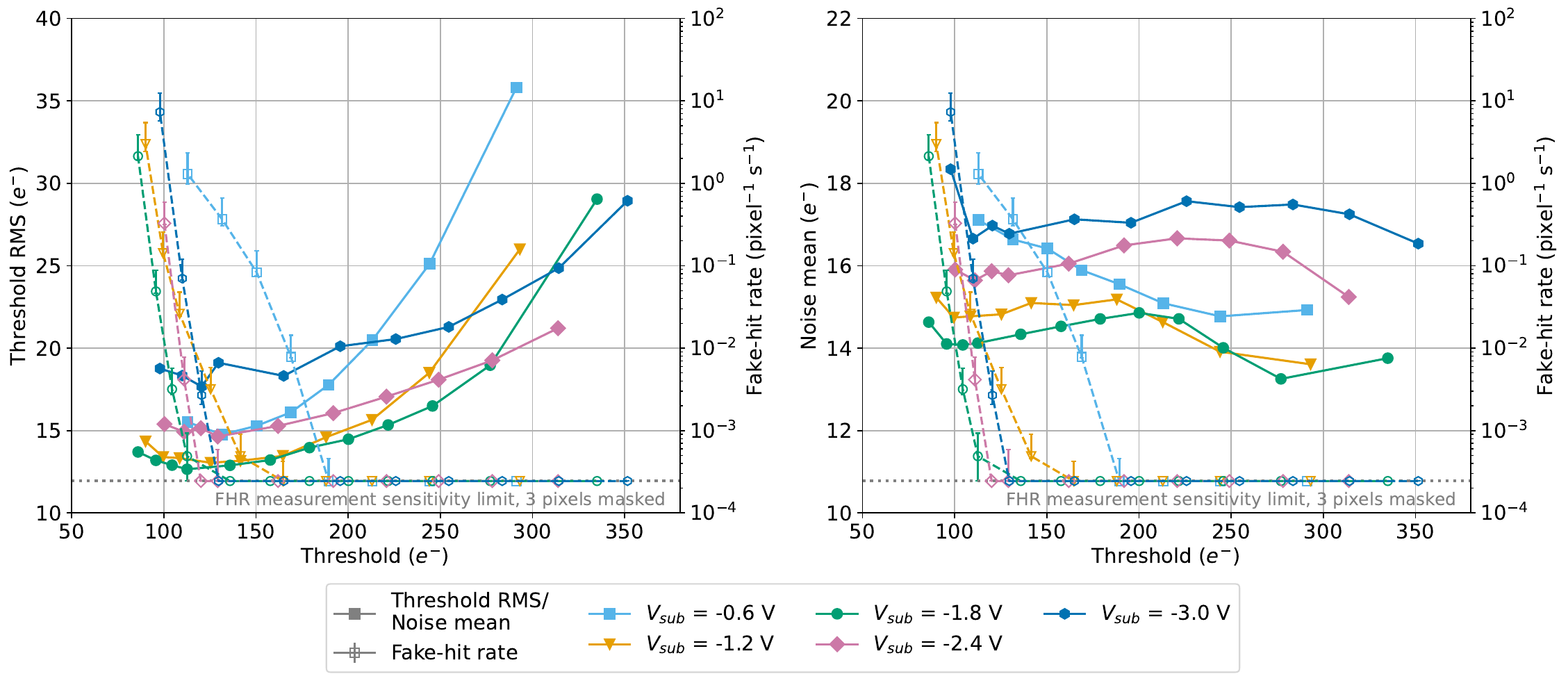}
    \caption{The threshold RMS and fake-hit rate vs. threshold (left) and the noise mean and fake-hit rate vs. threshold (right) at varying $\Vsub$ values at $\Ib=\qty{10}{\nA}$. At each $\Vsub$, the chip bias parameters were optimised.}
    \label{fig:pwr_thr_noise_fhr_vbb}
\end{figure}

An option to improve the performance at the very low power consumption operating points is to increase the reverse bias ($\Vsub$). This is shown in Fig.~\ref{fig:pwr_thr_noise_fhr_vbb}, which exhibits an improvement in terms of the onset of the FHR when increasing $\Vsub$ from $\qty{-1.2}{\volt}$ to $\qty{-3}{\volt}$. However, in terms of threshold RMS and noise mean, only $\qty{-1.8}{\volt}$ has marginally lower values, whereas for $\qty{-2.4}{\volt}$ and $\qty{-3}{\volt}$, the values are larger. These results indicate that at the lowest power setting, there are minimal improvements in the performance above $\Vsub=\qty{-1.2}{\volt}$.

\pagebreak

\subsection{Performance after ionising irradiation}
\label{sec:lab_tid}

\begin{figure}[!t]
    \centering
    \includegraphics[width=0.99\textwidth]
    {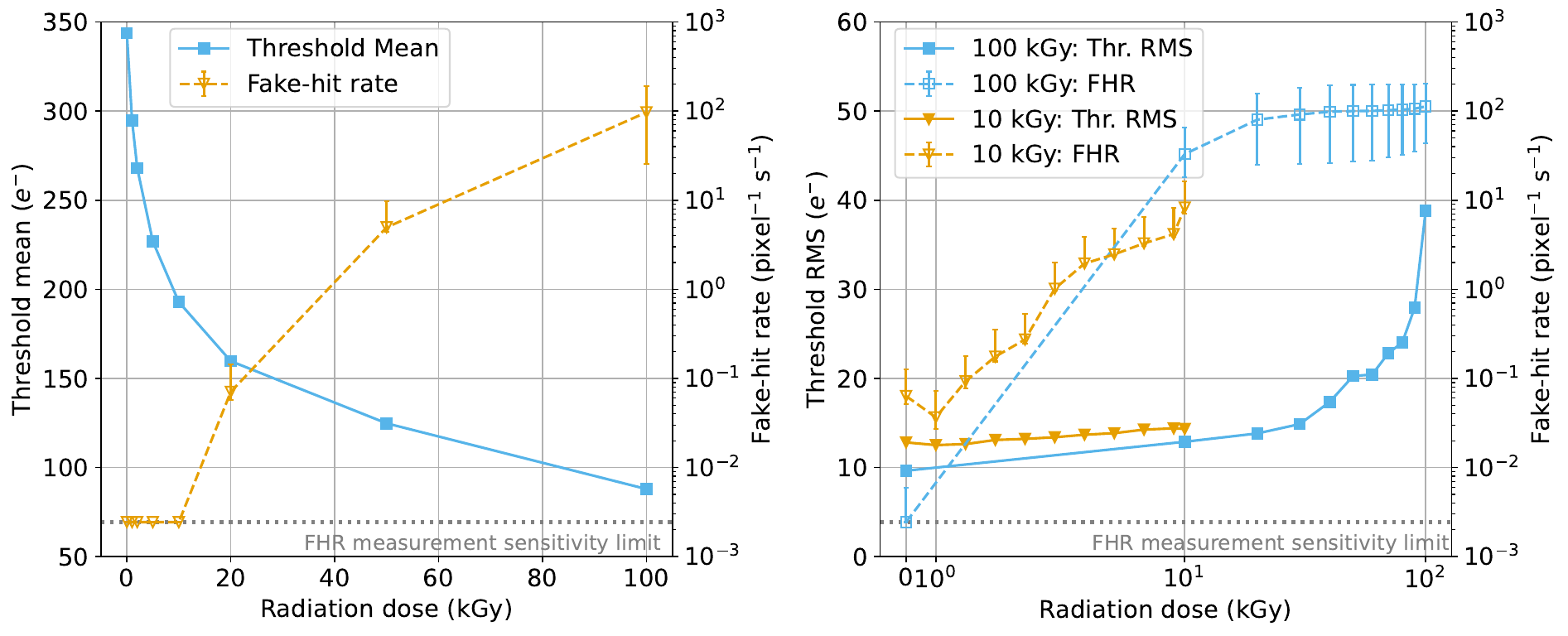}
    \caption{Left: the threshold mean and fake-hit rate vs. ionising radiation dose at a constant $\Vb$. Right: the threshold RMS and fake-hit rate vs. ionising radiation dose at a threshold tuned to \qty{125}{\ele} for two chips irradiated to different dose levels, \qty{10}{\kilo\gray} and \qty{100}{\kilo\gray}, at different rates.}
    \label{fig:tid_vs_dose}
\end{figure}

The evolution of the chip performance, in terms of threshold and fake-hit rate, was measured during the irradiation of several sensors using \qty{10}{\keV} X-rays from a tungsten target at CERN. The impact of the radiation dose at constant chip bias parameters is shown in Fig. \ref{fig:tid_vs_dose} (left), demonstrating that the threshold mean decreases and the fake-hit rate increases. This suggests that the radiation changes the chip's operating point, which can be accounted for by changing $\Vb$. To disentangle the effect of the changing threshold on the fake-hit rate, the performance was also measured while tuning the threshold to a constant value of \qty{125}{\ele} during the irradiation. The value of \qty{125}{\ele} was chosen so that the FHR would be visible at low doses, but not so high that the chip becomes inoperable at higher doses. The results for two chips irradiated to different dose levels, \qty{10}{\kilo\gray} and \qty{100}{\kilo\gray}, at different rates are shown in Fig. \ref{fig:tid_vs_dose} (right). It can be seen that the fake-hit rate at a constant threshold increases with radiation dose, increasing by approximately two orders of magnitude from \qtyrange{0}{10}{\kilo\gray}. For the threshold RMS, a slight increase is seen from \qtyrange{0}{10}{\kilo\gray}, but above this range, it increases rapidly.

\begin{figure}[!t]
    \centering
    \includegraphics[width=0.99\textwidth]
    {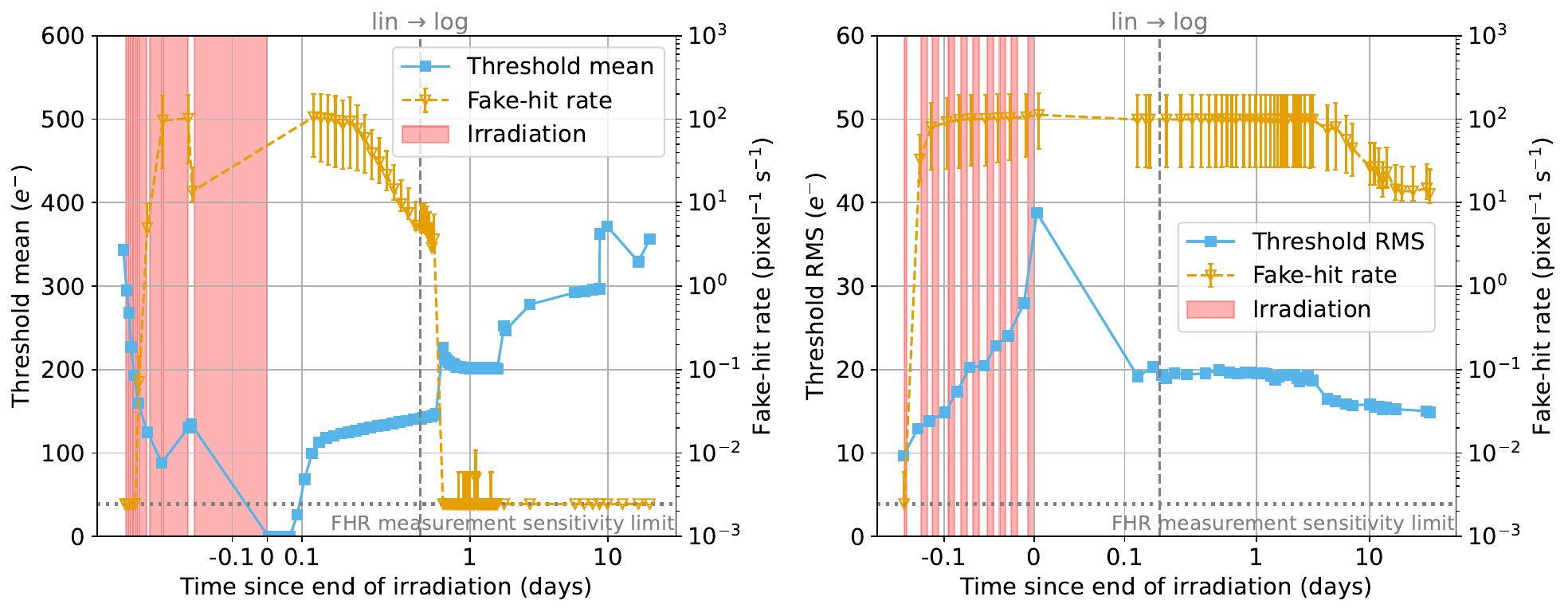}
    \caption{Left: the threshold mean and fake-hit rate vs. the time since the end of the irradiation at a constant $\Vb$ for a chip with a total dose of \qty{500}{\kilo\gray}. Right: the threshold RMS and fake-hit rate vs. the time since the end of the irradiation at a threshold tuned to \qty{125}{\ele} for a chip with a total dose of \qty{100}{\kilo\gray}. The red boxes indicate the period during which the sensor underwent irradiation.}
    \label{fig:tid_vs_annealing}
\end{figure}

The chip performance was also evaluated as a function of the time elapsed since the end of the irradiation. This annealing was performed at a constant temperature of \SIrange{21}{22}{\celsius} as the irradiated chips were stored and tested in a temperature-controlled laboratory. Figure \ref{fig:tid_vs_annealing} (left) shows the annealing effect on a chip with a total dose of \qty{500}{\kilo\gray} at constant chip bias parameters. After the irradiation, the fake-hit rate decreases by an order of magnitude within the first day. The sharp drop in the fake-hit rate is likely due to a power cycle of the chip that occurred during this time. The general trend of the threshold mean is that it increases with annealing time. Immediately after the irradiation, the points where the threshold is 0 indicate that the chip is inoperable. After this, the chip begins to recover and the threshold can be measured. Once again, the sharp fluctuations are likely caused by power cycles.

Figure \ref{fig:tid_vs_annealing} (right) shows the performance of a chip tuned to a threshold of \qty{125}{\ele} with a total dose of \qty{100}{\kilo\gray}. The threshold tuning was performed before each threshold measurement on a subset of pixels. Here, the fake-hit rate remains constant until three days after the irradiation, at which point the fake-hit rate starts to decrease. The same trend is also seen in the threshold RMS. These measurements have been performed on several different chips at doses of \qtylist[list-units=single]{10;100;500}{\kilo\gray}, and they all show similar trends.

\begin{figure}[!t]
    \centering
    \includegraphics[width=0.75\textwidth]
    {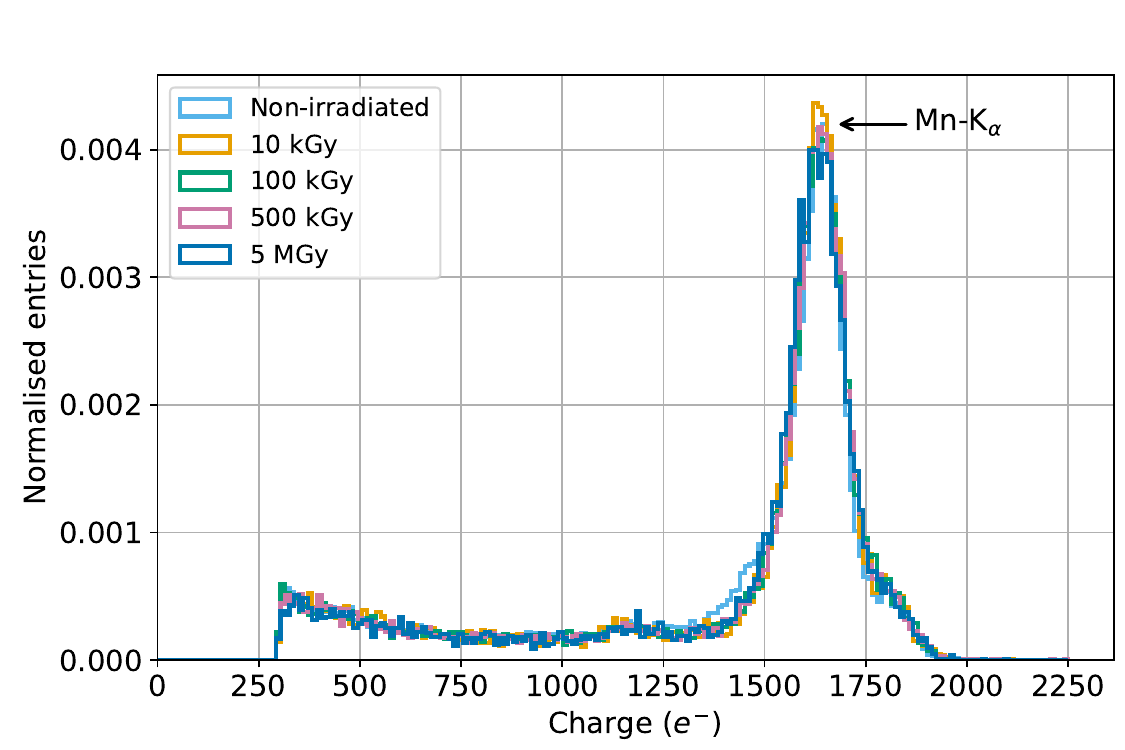}
    \caption{Measured $^{55}$Fe spectra of single pixel clusters for different levels of ionising irradiation: non-irradiated, 10, 100, \qty{500}{\kilo\gray} and \qty{5}{\mega\gray}.}
    \label{fig:fe55_tid_dose_comp}
\end{figure}

Another way to evaluate the chip performance after ionising irradiation is to compare the $^{55}$Fe spectra. To perform this test, X-ray emissions from an $^{55}$Fe source illuminating the top side of the chip at a distance of approximately 12~mm were measured. Three selections were applied to the data: keeping only single pixel clusters, removing all the edge pixels and removing the pixels whose hit rates are five standard deviations above the mean. A ToT calibration for each pixel is then applied to account for the variation in the pixel-to-pixel response, and a charge calibration to convert the measured ToT in mV to $\qty{}{\ele}$ is applied individually for each chip \cite{AGLIERIRINELLA2023168589}. Figure \ref{fig:fe55_tid_dose_comp} shows the spectra for a non-irradiated chip and chips irradiated to 10, 100, \qty{500}{\kilo\gray} and \qty{5}{\mega\gray}, where it can be seen that there is a negligible difference in the position, height and width of the $\text{Mn-K}_{\alpha}$ peak. This indicates, as expected, that ionising radiation has no impact on the charge collection up to dose values of \qty{5}{\mega\gray}.

\begin{figure}[!ht]
    \centering
    \includegraphics[width=0.725\textwidth]
    {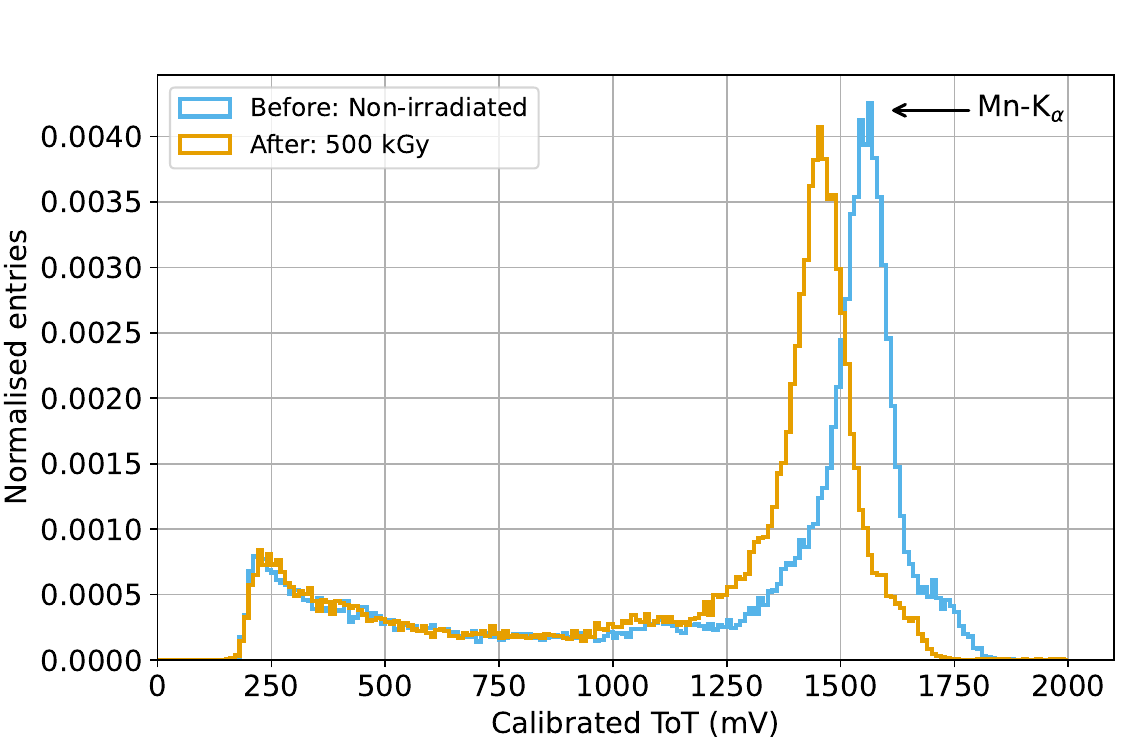}
    \caption{Measured $^{55}$Fe spectra of single pixel clusters of the same chip before and after an irradiation dose of \qty{500}{kGy} tuned to the same threshold (\qty{200}{\ele}).}
    \label{fig:fe55_tid_before_after_comp}
\end{figure}

However, the influence of ionising irradiation on the front-end electronics is seen by comparing the measured $^{55}$Fe spectrum of the same chip before and after irradiation tuned to the same threshold (\qty{200}{\ele}), as shown in Fig. \ref{fig:fe55_tid_before_after_comp}. A threshold of \qty{200}{\ele} was chosen to ensure that, after irradiation, the sensor’s FHR would remain low enough not to contribute a significant number of noise hits. The irradiation causes a shift of the $\text{Mn-K}_{\alpha}$ peak to lower ToT values, which points to an influence of the irradiation on the ToT calibration. Further evidence of this change is shown in Tab. \ref{tab:tid_charge_cal}, where the charge calibration (conversion factor of the ToT from \si{\mV} to \si{\ele}) is given before and after irradiation for different dose levels with 81~days of annealing measured at the same conditions. As the dose level increases, so does the difference in the charge calibration.

\begin{table}[!h]
\centering
\caption{Measured charge calibration factors before and after irradiation of the same sensor for chips at different levels of ionising irradiation.}
\begin{tabular}{l c c c}
    \toprule
    Dose (kGy) & Calibration before & Calibration after \\
    \midrule
     10  & 1.045$\pm$0.067 & 1.080$\pm$0.077 \\
     100 & 1.044$\pm$0.077 & 1.104$\pm$0.082 \\
     500 & 1.054$\pm$0.071 & 1.124$\pm$0.087 \\
    \bottomrule
\end{tabular}
\label{tab:tid_charge_cal}
\end{table}

\subsection{Fake-hit rate}
\label{sec:lab_fhr}

An important quantity for evaluating the sensor performance is the FHR, and how this quantity evolves with the chip biasing has already been measured~\cite{AGLIERIRINELLA2023168589}. The FHR is defined as the number of noise hits per pixel per second in the absence of external stimuli. For a MAPS, there are two main contributions to the total FHR: thermal noise and random telegraph noise (RTN). The contribution of thermal noise is random and can be modelled by a function that determines the probability that a pixel has a fake hit at a certain threshold. Assuming the thermal noise follows a zero-mean Gaussian distribution with standard deviation $n_{p}$, $\mathcal{N}(x|0,n_{p})$, the probability of a fake hit is given by

\begin{equation}
    \label{eq:pix_hit}
    f(t_{p},n_{p}) = \int_{t_{p}}^{\infty} \mathcal{N}(x|0,n_{p})\, dx = \frac{1}{2}\text{erfc}\left(\frac{t_{p}}{\sqrt{2}n_{p}}\right),
\end{equation}

\noindent where $t_{p}$ and $n_{p}$ are the threshold and noise of the pixel given by an s-curve measurement. To account for the pixel-to-pixel change in threshold and noise, the average fake-hit rate over the whole matrix is given by the expectation value of $f(t_{p},n_{p})$ \cite{Suljic:2303618}. Assuming the threshold and noise distributions are Gaussian, and partially solving gives:

\begin{equation}
    \label{eq:fhr}
    \overline{\text{FHR}}(\mu_{t},\sigma_{t},\mu_{n},\sigma_{n}) = \frac{1}{2}\int_{0}^{\infty} \text{erfc}\left(\frac{\mu_{t}}{\sqrt{2}\sqrt{n^2 + \sigma_{t}^2}}\right)\mathcal{N}(n|\mu_{n},\sigma_{n})dn,
\end{equation}

\noindent where $\mu_{t}$ and $\sigma_{t}$ are the mean chip threshold and threshold RMS, $\mu_{n}$ and $\sigma_{n}$ are the mean chip noise and noise RMS, and $\mathcal{N}(n|\mu_{n},\sigma_{n})$ is a normal distribution with mean $\mu_{n}$ and standard deviation $\sigma_{n}$. The remaining integral in Eq. \ref{eq:fhr} has to be solved numerically. There is no simple model for the RTN noise contribution to the FHR, so it is not included in the FHR model. 

\begin{figure}[t]
    \centering
    \includegraphics[width=0.68\textwidth]
    {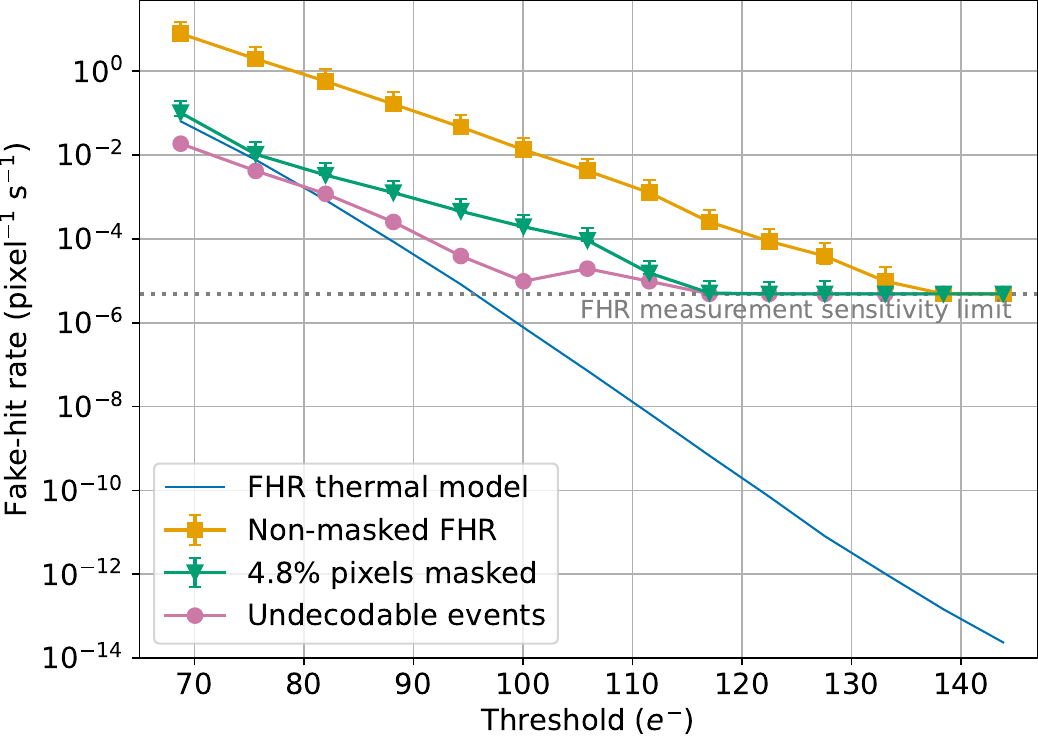}
    \caption{The measured fake-hit rate (line and dot markers) with different levels of masking, the contribution from undecodable events (orange) and the FHR model (blue line) as a function of threshold. Data were taken with \qty{5e6}{triggers} at a constant temperature of +\SI{20}{\celsius}.}
    \label{fig:fhr_with_model}
\end{figure}

Measurements of the FHR for non-irradiated DPTS chips at the `nominal' settings were performed as a function of the threshold with \qty{5e6}{triggers} at a constant +\SI{20}{\celsius}. A comparison of the measured results with the thermal FHR model described in Eq. \ref{eq:fhr} is shown in Fig. \ref{fig:fhr_with_model}. In total, three chips were measured and compared to the thermal model, and all show the same trends. To avoid selection bias and for visualisation purposes, the first chip that was measured was chosen to be presented in the paper. For the non-masked results, there is a large difference between the thermal model and the measurements, indicating that the thermal contribution is not the dominant noise source. 

A mask was then applied to the results by cutting pixels with a hit frequency higher than ten times the matrix mean frequency, resulting in \qty{4.8}{\percent} pixels masked. This masking reduces the FHR by around 2 orders of magnitude, indicating a large contribution from pixels that systematically give a hit, the so-called hot pixels. A hot pixel may occur due to defects within the pixel, causing its threshold to deviate significantly from the mean or resulting in high s-curve noise. The masked FHR results show agreement with the thermal FHR model at low thresholds (\qtyrange{70}{80}{\ele}), above which the measured values start to deviate from the model, signifying that other sources are significantly contributing to the masked FHR. One such contribution could be from the undecodable events (and thus unmaskable) that arise when hits clash on the CML output (cf. Sec. \ref{sec:dpts_chip}). Another source, particularly at higher thresholds, could be RTN. 

Overall, the main source of noise for the DPTS comes from the hot pixels. After masking these pixels, the contributions to the FHR are thermal noise at low thresholds, and the undecodable events and RTN at larger thresholds. A further validation of the thermal FHR model described here is needed and will be performed with the next generation of chip. This chip will not suffer from undecodable hits due to its different readout architecture, enabling a more thorough investigation of the agreement between the model and the FHR at lower thresholds

\begin{figure}[t]
    \centering
    \includegraphics[width=0.7\textwidth]
    {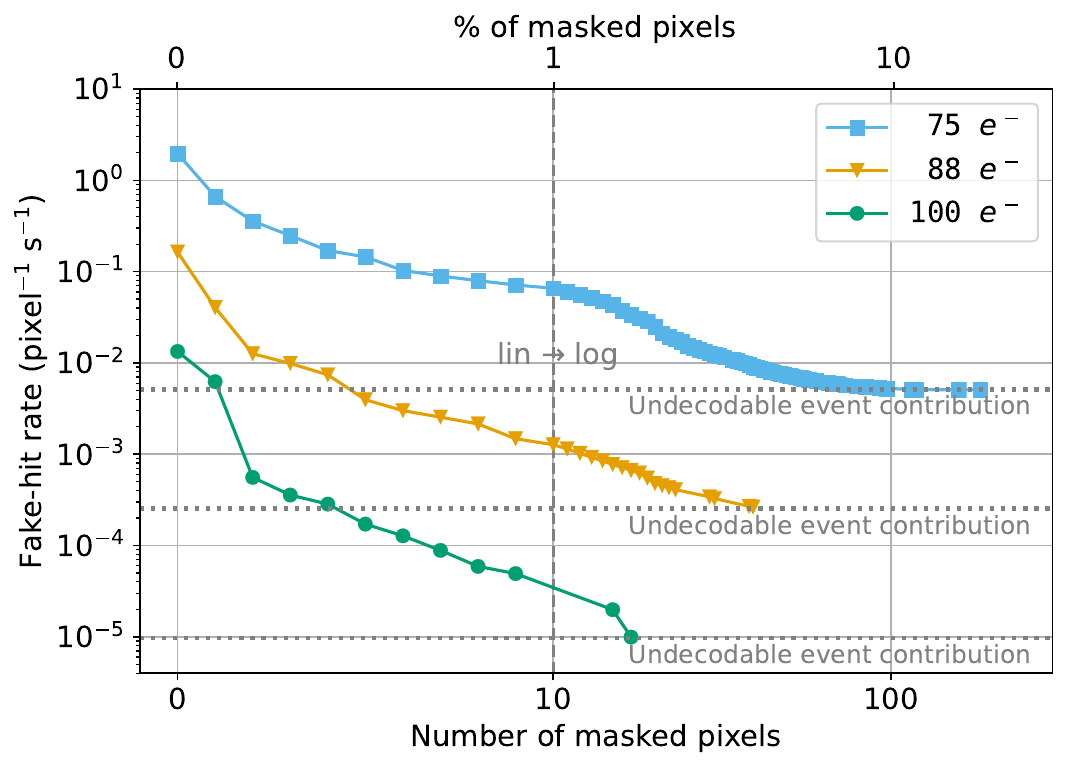}
    \caption{The measured fake-hit rate as a function of the number of pixels masked at three threshold values: \qtylist[list-units=single]{75;88;100}{\ele}. The $x$-axis initially uses a linear scale, but after reaching 10, it switches to a logarithmic scale.}
    \label{fig:fhr_vs_mask}
\end{figure}

Another important aspect to understand about FHR is how it varies with the number of pixels masked, as shown in Fig. \ref{fig:fhr_vs_mask} for three threshold values: \qtylist[list-units=single]{75;88;100}{\ele}. By masking 10 pixels (less than \qty{1}{\percent}), the FHR reduces by more than an order of magnitude, demonstrating that the FHR for this sensor is dominated by a few pixels that are consistently registering a hit. As the threshold increases, the reduction in FHR from masking 10 pixels becomes more pronounced. Masking additional pixels further reduces the FHR, albeit to a lesser extent, and eventually the FHR plateaus. This plateau occurs because the contribution to the FHR from undecodable events cannot be masked, indicating a non-negligible contribution from such hits. If these hot pixels can be identified and are few, they can be masked online, allowing the FHR to be lower and reducing the likelihood of having undecodable events.

\section{Measurements of fluorescence X-rays emitted from a target}
\label{sec:x-rays}

The DPTS sensor was characterised by fluorescence X-rays to evaluate the linearity of the front-end from 1.7 to 28.5~keV. The measurement campaign took place from May 2023 to June 2023 at the OptImaTo (Optimal Imaging and Tomography) laboratory~\cite{Trapani_2024}, located at Elettra Sincrotrone Trieste (Italy)~\cite{elettra}.

\subsection{Setup}
\label{sec:x-rays_setup}

A microfocus liquid metaljet X-ray source (Excillum D2+, 160 kV, Sweden) with a Galinstan alloy anode was used to produce the primary X-ray beam. A conical X-ray beam exits the source through a window with a 13-degree opening angle, and a copper collimator was used to further reduce the beam divergence. The energy spectrum of the beam was a continuum Bremsstrahlung, adjustable in the range 30-160~kV and contained characteristic lines of Gallium, Indium and Tin emissions (Tab.~\ref{tab:xray_beam_energy}).

\begin{table}[!b]
\centering
\caption{Energies of the characteristic peaks~\cite{xraydatabooklet} of the primary beam.}
\begin{tabular}{l c}
    \toprule
    Peak & Energy (keV) \\
    \midrule
     Ga K$\alpha$ & 9.2 \\
     In K$\alpha$ & 24.2 \\
     Sn K$\alpha$  & 25.2 \\
    \bottomrule
\end{tabular}
\label{tab:xray_beam_energy}
\end{table}

\pagebreak

\begin{figure}[!t]
  \centering
  \includegraphics[width=0.9\textwidth]{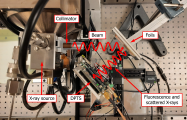}
  \caption{A picture of the setup used to measure fluorescence X-rays emitted from either a Manganese or tin foil.}
  \label{fig:x-raySetup}
\end{figure}

Manganese and tin foils were used as targets and placed in the beam with an angle of 45$^{\circ}$ relative to the beam direction. The characteristic emissions of manganese and tin are shown in Tab.~\ref{tab:mn_sn_emissions}. The DPTS was placed parallel to the target at approximately \qty{10}{\cm}, with the top side facing the incoming fluorescence emissions. This position was chosen to remove the detector from the conical beam and reduce unwanted hits from the primary X-rays. However, hits from Compton scattering of the primary beam on the target are still possible. A picture of the setup is shown in Fig. \ref{fig:x-raySetup}.

\begin{table}[t]
\centering
\caption{Energies of the characteristic peaks~\cite{xraydatabooklet} of the manganese and tin targets.}
\begin{tabular}{l c}
    \toprule
    Peak & Energy (keV) \\
    \midrule
     Mn K$\alpha$ & 5.9 \\
     Mn K$\beta$  & 6.5 \\
     Sn K$\alpha$ & 25.3 \\
     Sn K$\beta$  & 28.5 \\
     Sn L$\alpha$ & 3.4 \\
    \bottomrule
\end{tabular}
\label{tab:mn_sn_emissions}
\end{table}

The custom-made setup presented in Sec.~\ref{sec:lab_daq} was used to operate the DPTS. The chip biases were set to the nominal values except for $\Vb$, which was set to 250~mV to tune the threshold to 150~$e^{-}$ to reduce the chance of a noise hit.

\subsection{Results}
\label{sec:x-rays_results}

The data were analysed as described in Sec.~\ref{sec:lab_tid} for the $^{55}$Fe spectrum. An additional selection that removes pixels with bad position encoding has been applied to the data.

\pagebreak

\begin{figure}[!t]
  \centering
  \includegraphics[width=0.87\textwidth]{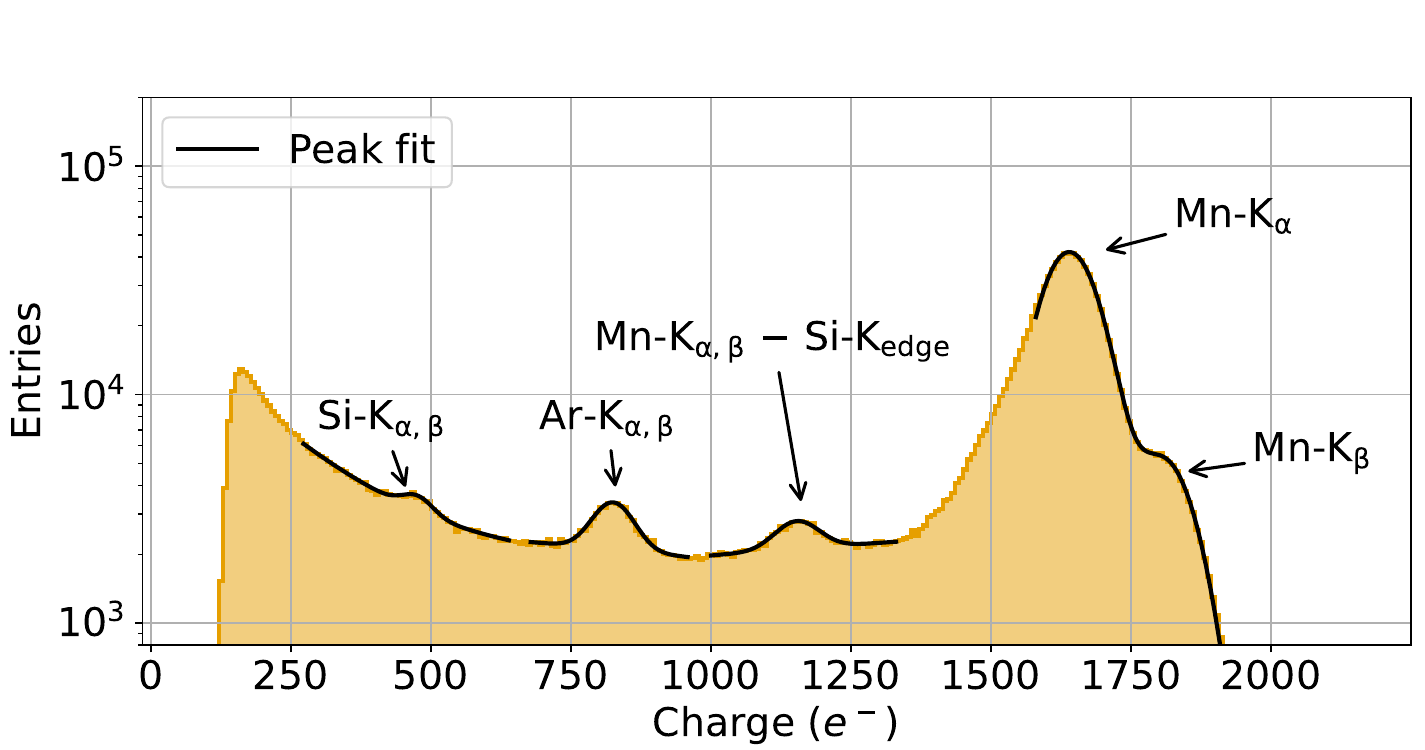}
  \caption{Measured spectrum of fluorescence X-rays emitted from a manganese target. The sensor was able to resolve the manganese characteristic emission peaks ($\text{Mn-K}_{\alpha}$ and $\text{Mn-K}_{\beta}$), the $\text{Mn-K}_{\alpha,\beta}$ escape peak ($\text{Mn-K}_{\alpha,\beta} - \text{Si-K}_{\mathrm{edge}}$), the argon characteristic emission peak ($\text{Ar-K}_{\alpha,\beta}$) and the silicon fluorescence peak ($\text{Si-K}_{\alpha,\beta}$).}
  \label{fig:manganese_spectrum}
\end{figure}

Figure~\ref{fig:manganese_spectrum} shows the measured spectrum of fluorescence X-rays emitted from a manganese target. Five emission peaks are resolved in the spectrum (Fig.~\ref{fig:manganese_spectrum}): the Mn-K$_{\alpha}$ and Mn-K$_{\beta}$ characteristic emission peaks, the $\text{Mn-K}_{\alpha,\beta}$ escape peak, the argon characteristic emission peak $\text{Ar-K}_{\alpha,\beta}$ and the silicon fluorescence peak $\text{Si-K}_{\alpha,\beta}$. The Mn-K$_{\alpha}$ and Mn-K$_{\beta}$ peaks are fitted with a sum of two Gaussians. The escape and the $\text{Ar-K}_{\alpha,\beta}$ peaks are fitted with a Gaussian and a linear fit is used to reproduce the background. The silicon fluorescence peak is fitted with a Gaussian and an exponential fit is used to reproduce the background.

\begin{figure}[!t]
  \centering
  \includegraphics[width=0.99\textwidth]{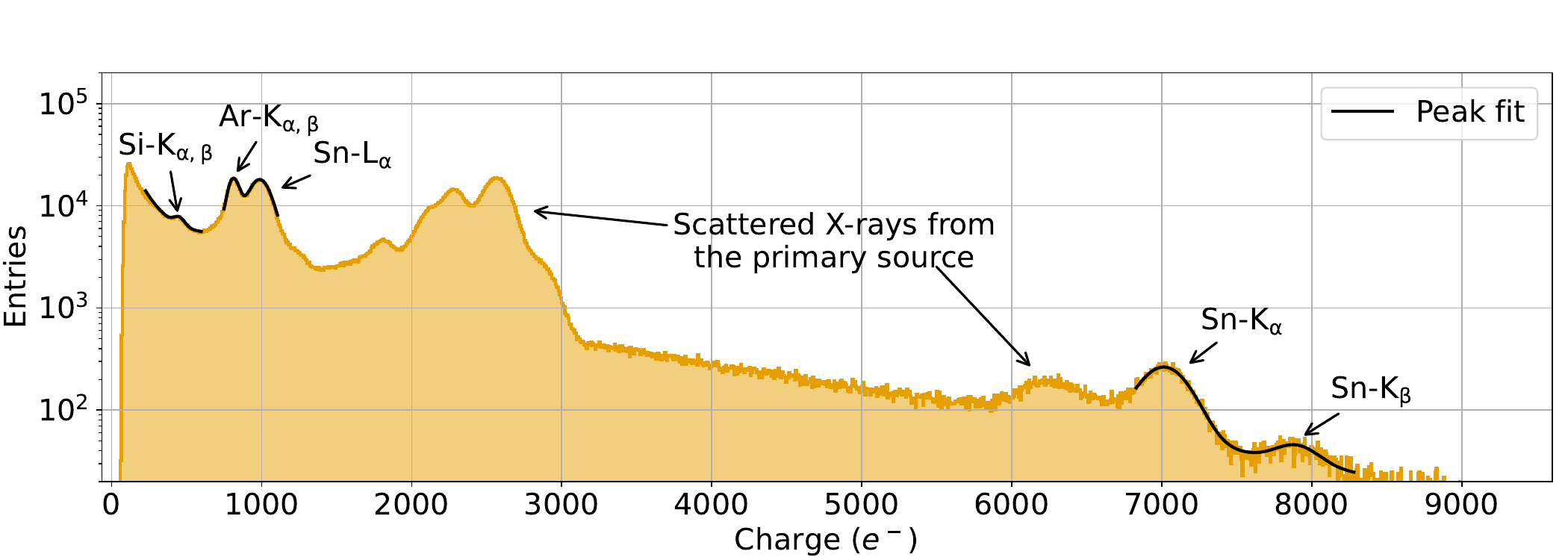}
  \caption{Measured spectrum of fluorescence X-rays emitted from a tin target. The sensor was able to resolve the Sn characteristic emission peaks ($\text{Sn-K}_{\alpha}$, $\text{Sn-K}_{\beta}$, $\text{Sn-L}_{\alpha}$), the argon characteristic emission peak ($\text{Ar-K}_{\alpha,\beta}$) and the silicon fluorescence peak ($\text{Si-K}_{\alpha,\beta}$).}
  \label{fig:tin_spectrum}
\end{figure}

Figure~\ref{fig:tin_spectrum} shows the measured spectrum of fluorescence X-rays emitted from a tin target. The emission peaks that are resolved in the spectrum (Fig.~\ref{fig:tin_spectrum}) are the Sn-K$_{\alpha}$, Sn-K$_{\beta}$ and Sn-L$_{\alpha}$ characteristic emission peaks, the argon characteristic emission peak $\text{Ar-K}_{\alpha,\beta}$, the silicon fluorescence peak $\text{Si-K}_{\alpha,\beta}$ and the peaks originating from scattered X-rays from the primary source. The Sn-K$_{\alpha}$ peak, the Sn-K$_{\beta}$ peak and the closest peak from scattered X-rays are fitted with a sum of three Gaussians and an exponential fit is used to reproduce the background. In Fig.~\ref{fig:tin_spectrum} the fit function is shown only for the Sn-K$_{\alpha}$ and the Sn-K$_{\beta}$ peaks. The $\text{Ar-K}_{\alpha,\beta}$ and the Sn-L$_{\alpha}$ peaks are fitted with a sum of two Gaussians and a linear background fit. The silicon fluorescence peak is fitted with a Gaussian, and an exponential fit is used to reproduce the background.

\begin{figure}[!t]
  \centering
  \includegraphics[width=0.68\textwidth=false]{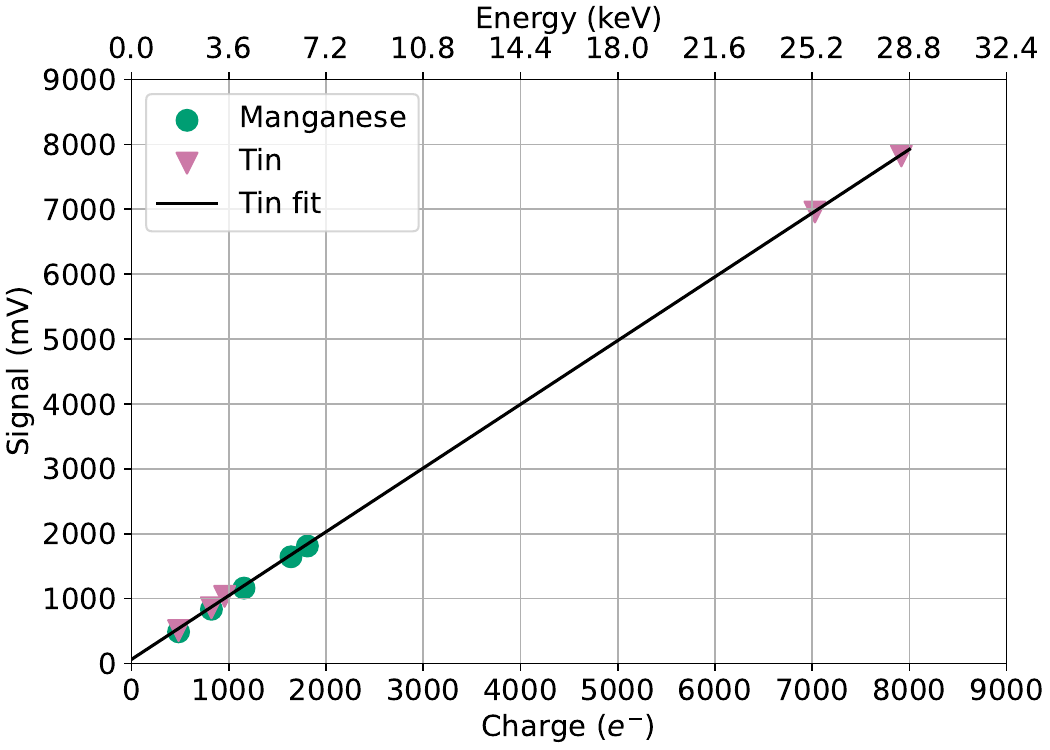}
  \caption{The measured peak values as a function of the literature peak values for both the manganese and tin targets. A linear function is fit to the peaks of the tin target.}
  \label{fig:linearity_response}
\end{figure}

The values of the peaks measured for both targets are shown in Fig.~\ref{fig:linearity_response}. The data from the tin target are fitted with a linear function. 
The linearity is preserved up to 28.5~keV, the energy of the tin K$_{\beta}$ emission.

\section{Measurements with ionising particle beams}
Following the laboratory results previously shown in Sec.~\ref{sec:lab_meas}, the DPTS performance was also measured with ionising particle beams to further investigate the behaviour at low-power consumption settings and after ionising irradiation. Additionally, an investigation into the performance with inclined tracks was carried out. These studies took place from July 2022 to May 2023 at two test beam facilities: CERN PS and DESY~\cite{desyII}, providing \qty{10}{\giga\electronvolt\per\textit{\text{c}}} positive hadrons and \qty{5.4}{\giga\electronvolt\per\textit{\text{c}}} electrons, respectively.

\subsection{Setup}
\label{sec:tb_setup}

\begin{figure}[!htb]
	\centering
    \input{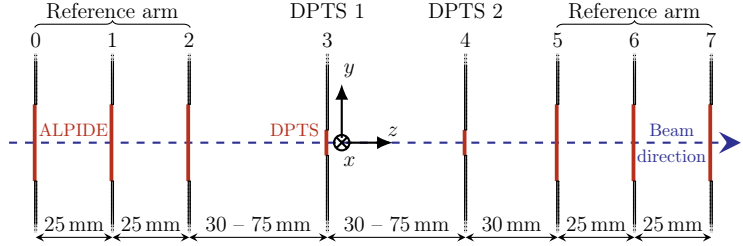}
    \caption{A sketch of the beam telescope (not to scale) used to perform measurements with ionising particle beams. The setup consists of 6 ALPIDE reference planes and two DPTS sensors in the centre. The distances on either side of the DUT were increased to \qty{75}{\mm} to accommodate the rotational stage for the inclined track measurements.}
    \label{fig:tb_setup}
\end{figure}

The setup consists of a beam telescope with six reference planes, each equipped with an ALPIDE chip. These planes provide space points for track reconstruction for two DPTS sensors installed at the centre of the telescope with three reference planes on either side, Fig.~\ref{fig:tb_setup}. One of the DPTSs was the device under test (DUT), while the other was used as a trigger. This trigger DPTS was installed on a micro-positioning stage to control the $x$ and $y$ movement so that an overlap with the DUT was made. The data acquisition was based on the EUDAQ2 framework~\cite{eudaq}. During the measurements with inclined tracks, the DUT's rotation around the $y$-axis was controlled with a rotational stage with an angle of \qty{0}{\degree} being the ``nominal'' position, i.e. the DUT is perpendicular to the beam. To account for the rotational stage, the distances to the adjacent planes on either side of the DUT were increased. For the measurements performed during the power-consumption and ionising-radiation studies, an aluminium jig with a \qty{1}{\mm} hole aligned to the position of the chip was used to cool and keep the DUT at a constant temperature of \SI{+20}{\celsius}.

\subsection{Analysis tools and methods}
\label{sec:tb_analysis}

The Corryvreckan test beam data reconstruction framework~\cite{corryvreckan} was used to process and analyse the data. This was done by reconstructing tracks from clusters on the reference planes with the General Broken Lines model~\cite{gbl_formalism} and interpolating the track position on the DUT(s). To ensure a high quality of the data sample, several track and event selection criteria were applied: only one track per event, a reduced track $\chi^{2}<3$, each track must have one hit on each reference plane, and tracks must not intersect the DUT within two pixel pitches from the sensor edge. Fake hits were kept to a minimum by masking pixels on the reference planes with too large a hit rate (more than \num{1000}~times the average; $<1$~pixel per plane were affected). For the DUT, the probability of associating a fake (noise) hit to a track was minimised by requiring the response to be within \SI{3}{\micro\second} of the trigger signal.

In the detection efficiency analysis, the non-decodable events arising from readout signal collisions on the CML lines were associated with the pixel (15, 15). These events are considered as real hits with an undetermined position due to the negligible likelihood that two fake hits coincide in time and result in a non-decodable event. As a result, a loose spatial selection is applied to ensure the efficiency is not underestimated by not including the non-decodable hits. This selection is a circular acceptance window of \qty{480}{\um} in radius used to associate clusters on the DUT to tracks.

In the timing resolution analysis, a spatial acceptance window selection with a radius of \qty{45}{\um} is used, but no timing selection is applied. Without an external trigger, the trigger DPTS was used as a timing reference and its chip biases were kept at the ``nominal'' settings throughout the measurements. Two corrections were applied to the measured DUT signal time accounting for the column delay and the time walk\footnote{These corrections were not possible with the trigger DUT since it was used as a timing reference.}. The time residuals between the DUT and the trigger DPTSs were then fit with a Gaussian to obtain the standard deviation $\sigma$. To estimate the timing resolution of the trigger DPTS, three factors need to be considered: the trigger DPTS was used as a reference, the timing resolution does not depend on chip biases other than $\Ib$, and there is a similar timing resolution performance for the DUT and trigger DPTSs. Therefore, the trigger timing resolution at the ``nominal'' $\Ib=\qty{100}{\nA}$ was estimated to be $\sigma/\sqrt{2}$, resulting in a value of $\sigma_{\text{TRG}} = \qty{28.3}{\ns}$. Finally, the DUT timing resolution for other $\Ib$ values was calculated using $\sigma_{\text{DUT}} = \sqrt{\sigma^{2} - \sigma_{\text{TRG}}^{2}}$.

\subsection{Performance at low power consumption}
\label{sec:tb_results_power}

Although previous performance results of the DPTS sensor are reported at a comparably high conservative pixel current of $\Ib=\qty{100}{\nA}$~\cite{AGLIERIRINELLA2023168589} and thus a high power consumption setting (cf. Sec.~\ref{sec:dpts_chip}), the sensor performance at low power consumption is of special interest for air cooling solutions in a future application of the technology.

\begin{figure}[!t]
    \centering
    \includegraphics[width=0.99\textwidth]
    {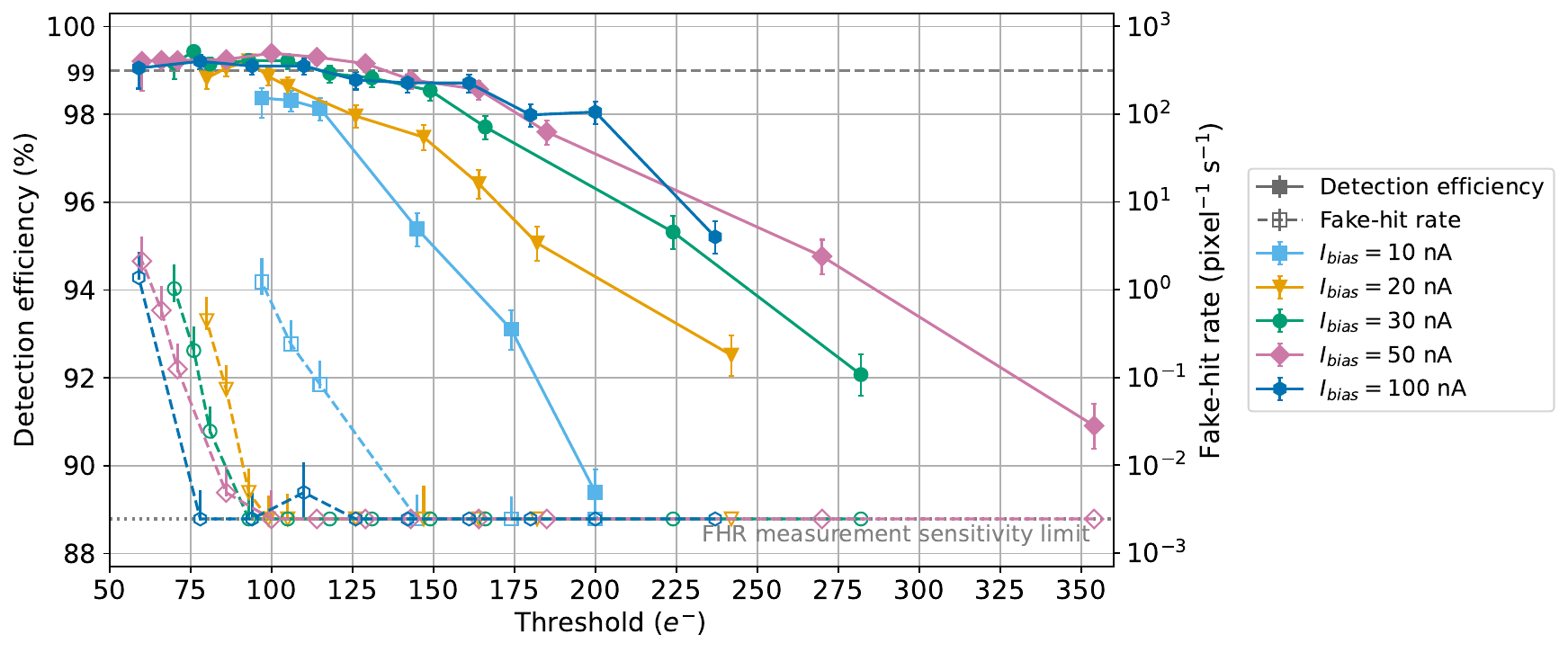}
    \caption{Detection efficiency (filled symbols, solid lines) and fake-hit rate (open symbols, dashed lines) as a function of the average threshold for a non-irradiated chip at different power consumption settings ($\Ib$).}
    \label{fig:eff_thr_power_consumption_non_irrad}
\end{figure}

Figure~\ref{fig:eff_thr_power_consumption_non_irrad} shows the dependence of the detection efficiency and the fake-hit rate, measured in situ, on the mean chip threshold achieved by supplying different voltages for $\Vb$ at various power consumption settings ranging from $\Ib=\qty{100}{\nano\ampere}$ to the lowest possible $\Ib=\qty{10}{\nano\ampere}$. The nominal reverse bias voltage of \qty{-1.2}{\volt} is supplied and the remaining biasing parameters are individually optimised for each $\Ib$ setting. The three noisiest pixels, determined by their accumulated hit occupancy during data analysis, are masked. Starting at \qty{100}{\nano\ampere} and reducing $\Ib$, it appears that the threshold margin in which efficiency remains above \qty{99}{\percent} is reduced and the onset of the FHR occurs at higher thresholds. For the ITS3, a FHR below \qty{0.01}{\per\pixel\per\s} is required. Thus, an $\Ib$ current of at least \qty{30}{\nA} is needed to be below this and reach \qty{99}{\percent} efficiency. This results in a pixel matrix power consumption of \qty{15.9}{\milli\watt\per\cm^2} for the DPTS but for the ITS3 sensor, with a larger pitch of $20.8\times22.8$~\si{\um}, this reduces to \qty{7.6}{\milli\watt\per\cm^2}, which is below the required target of \qty{15}{\milli\watt\per\cm^2} \cite{The:2890181}. As with the laboratory measurements in Sec.~\ref{sec:lab_power}, the chip performance drops considerably once at \qty{20}{\nano\ampere} and particularly for \qty{10}{\nano\ampere} where an efficiency of \qty{99}{\percent} is not reached.

\begin{figure}[!t]
    \centering
    \includegraphics[width=0.99\textwidth]
    {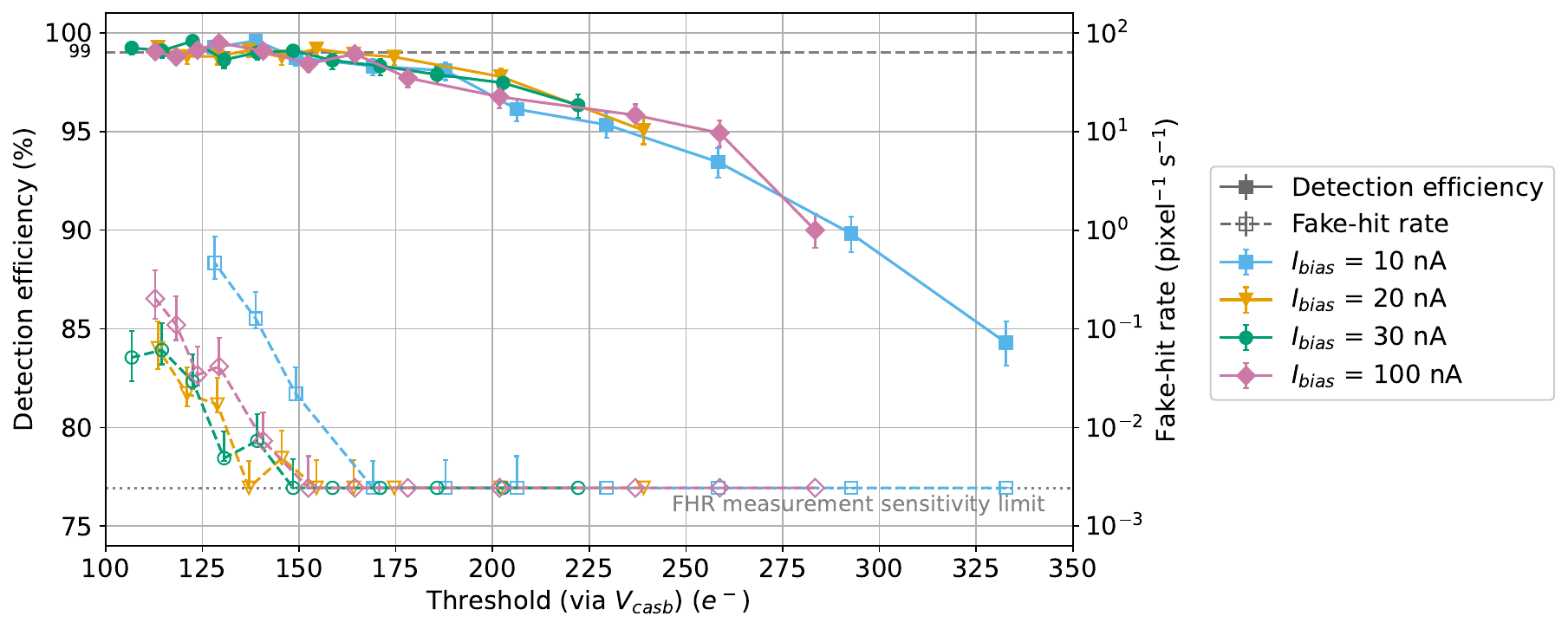}
    \caption{Detection efficiency (filled symbols, solid lines) and fake-hit rate (open symbols, dashed lines) as a function of the average threshold for a proton irradiated chip at different power consumption settings ($\Ib$) at $\Vsub=\SI{-1.2}{\volt}$. Three noisy pixels have been masked.}
    \label{fig:eff_thr_power_consumption_irrad_comp}
\end{figure}

To study the effect of radiation, the results are shown in Fig. \ref{fig:eff_thr_power_consumption_irrad_comp} for a proton irradiated sensor having received a combination of an ionising dose of \qty{10}{\kilo\gray} and a non-ionising dose of \qty{1e13}{\niel} (satisfying the requirements of ITS3~\cite{The:2890181}). Despite the shift in operational range, the irradiated chip shows no performance degradation in terms of efficiency and FHR performance compared to that of the non-irradiated chip. The irradiated chip can reach an efficiency of \qty{99}{\percent} at $\Ib=\qty{10}{\nano\ampere}$; however, this is likely attributed more to chip-to-chip variations than the irradiation.

The influence of the reverse bias voltage at $\Ib=\qty{10}{\nano\ampere}$ was also investigated for the non-irradiated chip and found a marginal improvement in performance when increasing $\Vsub$, but not significantly enough to justify having the chip operate in a very low-power consumption regime.

\begin{figure}[!t]
    \centering
    \includegraphics[width=0.8\textwidth]
    {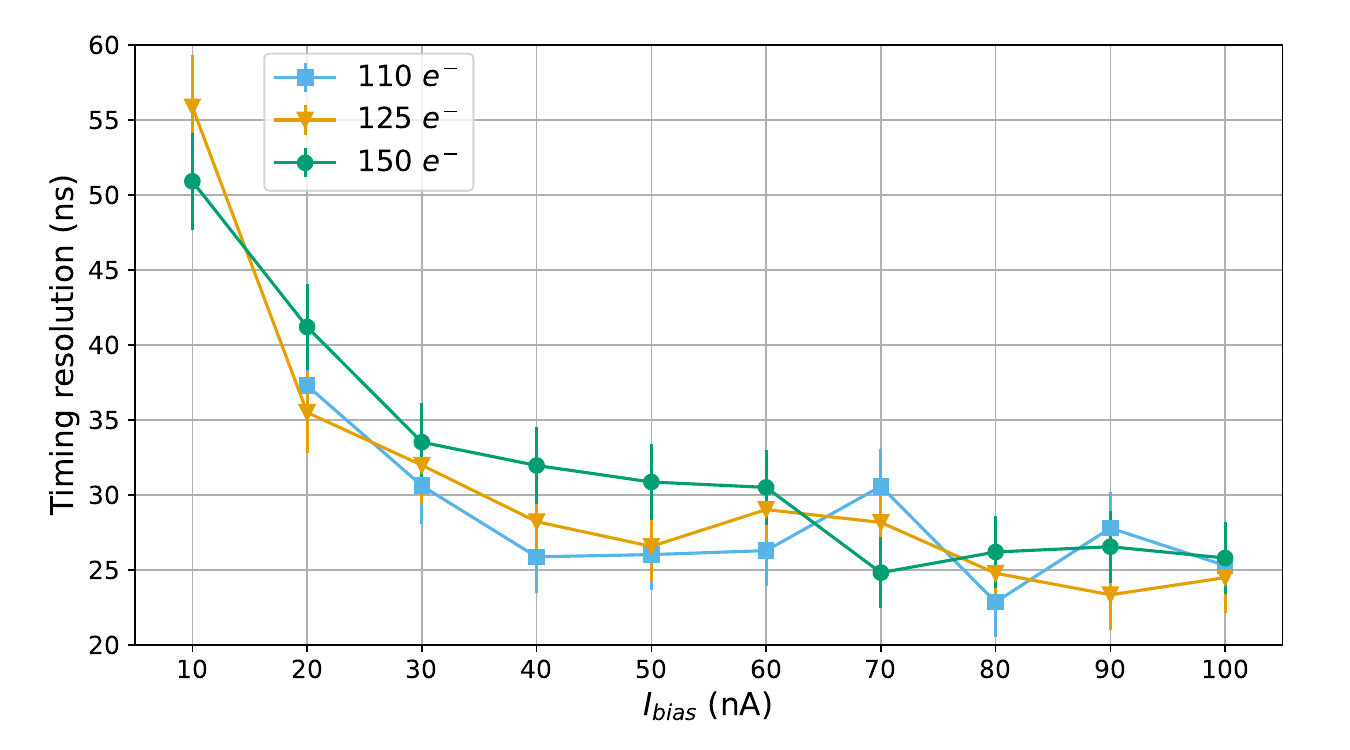}
    \caption{The timing resolution as a function of $\Ib$ at constant threshold values for a non-irradiated chip.}
    \label{fig:timing_vs_ibias}
\end{figure}

Further investigation of the chip performance at low power consumption was carried out by measuring the timing resolution. This is because $\Ib$ directly influences the timing performance of the DPTS. The impact of varying $\Ib$ on the timing resolution at a constant tuned threshold is shown in Fig. \ref{fig:timing_vs_ibias} where it can be seen that decreasing $\Ib$ from the ``nominal'' value of $\qty{100}{\nA}$ increases the timing resolution\footnote{The previous result of \qty{6.3}{\ns} \cite{AGLIERIRINELLA2023168589} was not replicated due to the different trigger system and no timewalk correction applied to the trigger DPTS.}. This trend becomes more significant once the $\Ib$ value drops below \qty{40}{\nA}, at which point, the time resolution increases more rapidly. For the three threshold values measured (\qtylist{110;125;150}{\ele}), there is a negligible influence of the threshold on the timing resolution.

\begin{figure}[!t]
    \centering
    \includegraphics[width=0.8\textwidth]
    {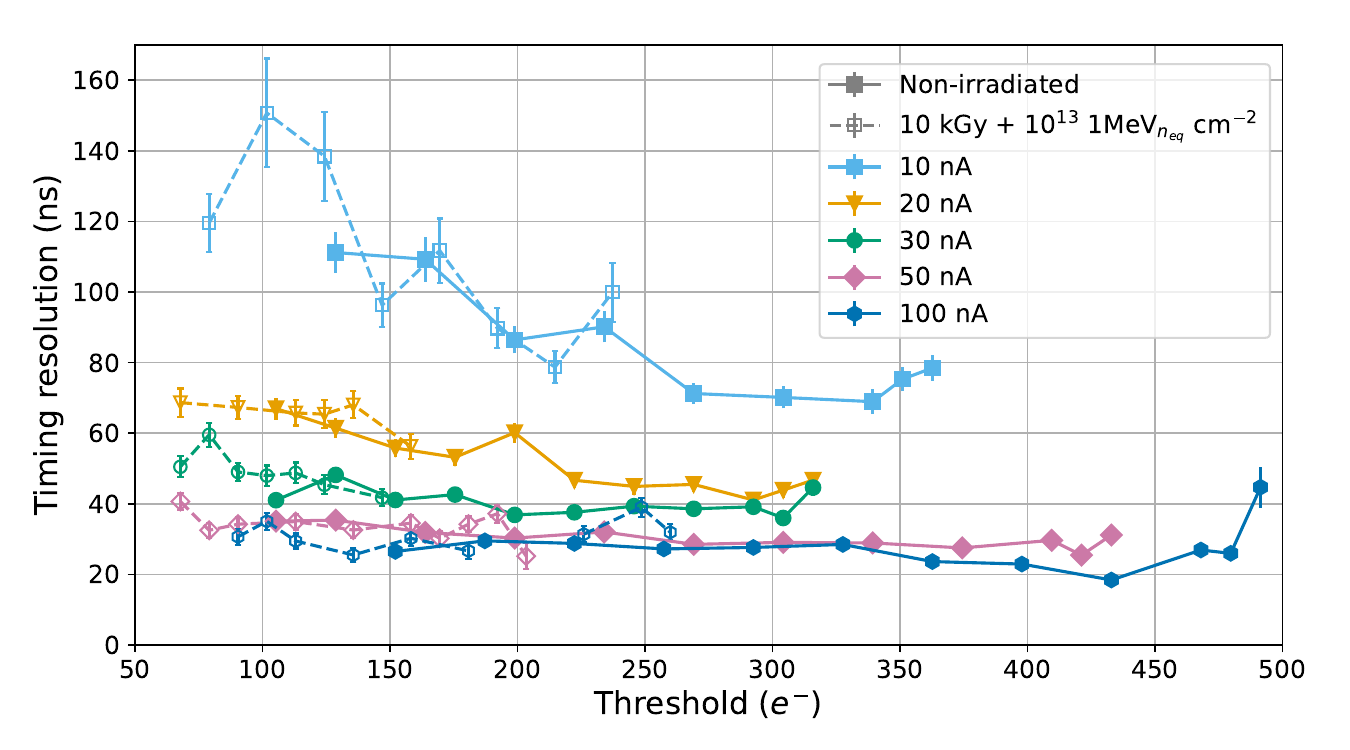}
    \caption{The timing resolution as a function of the threshold at varying $\Ib$ values for a non-irradiated chip (filled symbols, solid lines) and a proton irradiated chip (open symbols, dashed lines).}
    \label{fig:timing_vs_thr_ibias}
\end{figure}

Measurements of timing resolution were also carried out at chip bias settings that were optimised for each $\Ib$ to obtain the best performance. A comparison of the timing resolution as a function of the threshold between a non-irradiated chip and a chip irradiated to the expected ITS3 levels (\qty{10}{\kilo\gray} and \qty{1e13}{\niel}) is shown in Fig. \ref{fig:timing_vs_thr_ibias}. The influence of $\Ib$ is evident in both chips. The effect of the threshold on the timing resolution is negligible for $\Ib$ values $>\qty{30}{\nA}$. For lower values of $\Ib$, the timing resolution increases for lower thresholds. The cause of this trend could be that, at the lower values of $\Ib$, the operating point of the chip becomes less stable and affects the chip's performance. At the same threshold and $\Ib$ values, the non-irradiated chip and the irradiated chip show the same timing resolution, indicating that this level of irradiation has no impact on the timing performance.

\subsection{Performance after ionising irradiation at nominal power consumption}
\label{sec:tb_results_tid}

\begin{figure}[!t]
    \centering
    \includegraphics[width=0.99\textwidth]
    {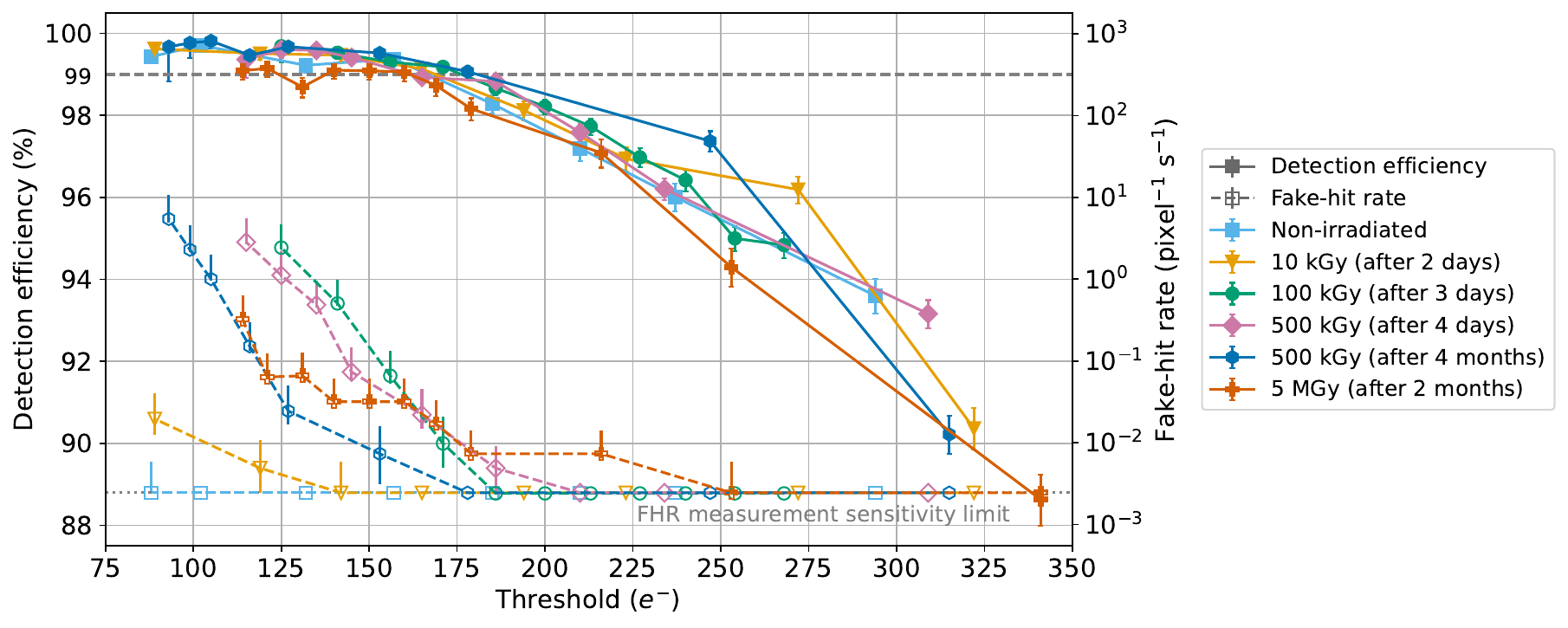}
    \caption{Detection efficiency (filled symbols, solid lines) and fake-hit rate (open symbols, dashed lines) as a function of the average threshold for chips irradiated to various ionising doses and annealing times. The chips were measured at ``nominal'' setting except for $\Vsub=\qty{-3.0}{\volt}$. Four pixels have been masked only for the \qty{5}{\mega\gray} chip.}
    \label{fig:eff_thr_tid}
\end{figure}

Previously, it was shown that the DPTS sensor is radiation hard up to ionising doses of \qty{100}{\kilo\gray}~\cite{AGLIERIRINELLA2023168589}. This radiation hardness was further tested by first irradiating the chips at CERN (cf. Sec. \ref{sec:lab_tid}) at dose levels up to \qty{5}{\mega\gray}, then measuring the performance. Results of these irradiated chips are shown in Fig.~\ref{fig:eff_thr_tid}, where the detection efficiency and fake-hit rate are shown as functions of the average threshold at ``nominal'' setting, except for $\Vsub=\qty{-3.0}{\volt}$ where the best performance in terms of efficiency and FHR is seen~\cite{AGLIERIRINELLA2023168589}. It can be seen that a detection efficiency of \qty{99}{\percent} is reachable for all dose levels and that with increasing dose, the onset of the FHR occurs at higher thresholds. For the chip irradiated to \qty{500}{\kilo\gray}, the impact of annealing on the chip is also seen, showing that the onset of the FHR occurs at lower thresholds when the chip is annealed for four months compared to 4 days. This means there is a larger operational region where the detection efficiency is above \qty{99}{\percent} and the FHR is below \qty{0.01}{\per\pixel\per\s}.

\subsection{Inclined tracks}
\label{sec:tb_results_inclined}

\begin{figure}[!t]
    \centering
    \includegraphics[width=0.99\textwidth]
    {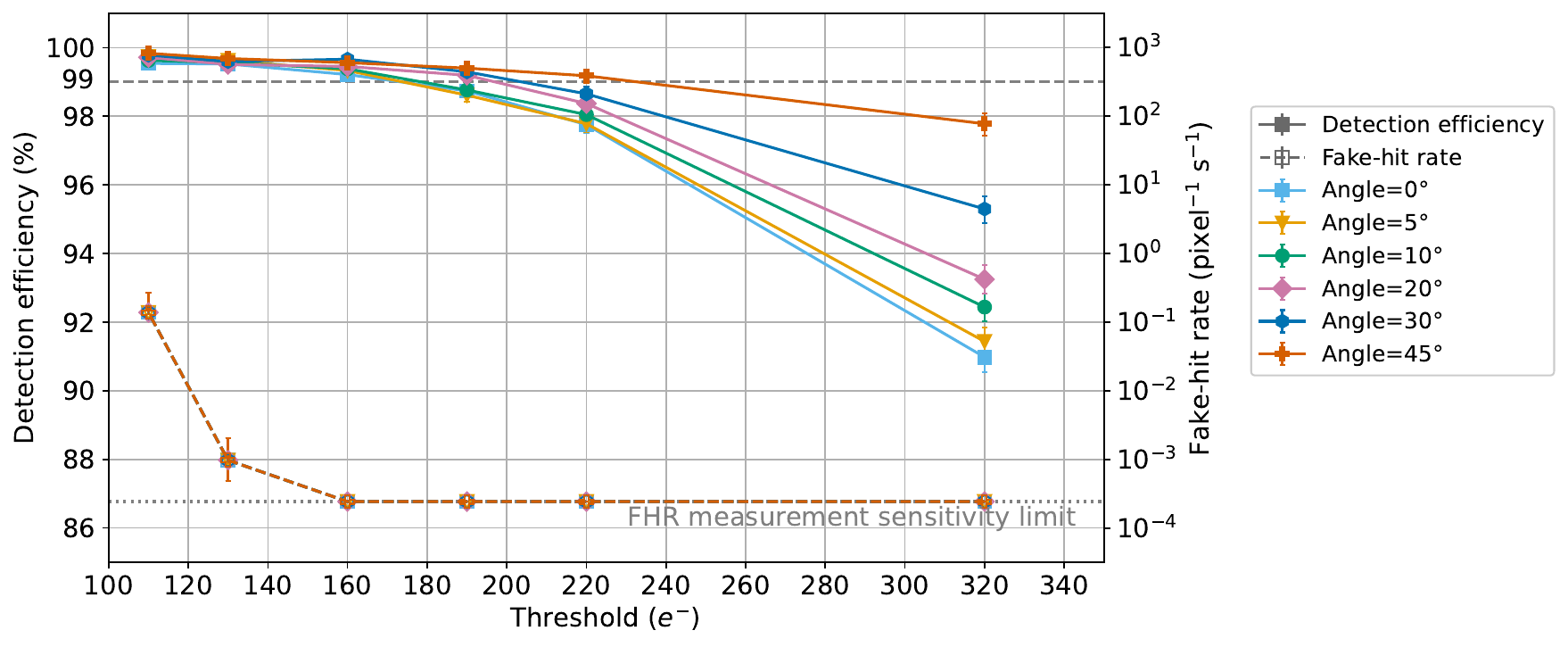}
    \caption{Detection efficiency (filled symbols, solid lines) and fake-hit rate (open symbols, dashed lines) as a function of the average threshold for chips with different track incident angles.}
    \label{fig:eff_vs_thr_angle}
\end{figure}

\begin{figure}[!t]
    \centering
    \includegraphics[width=0.99\textwidth]
    {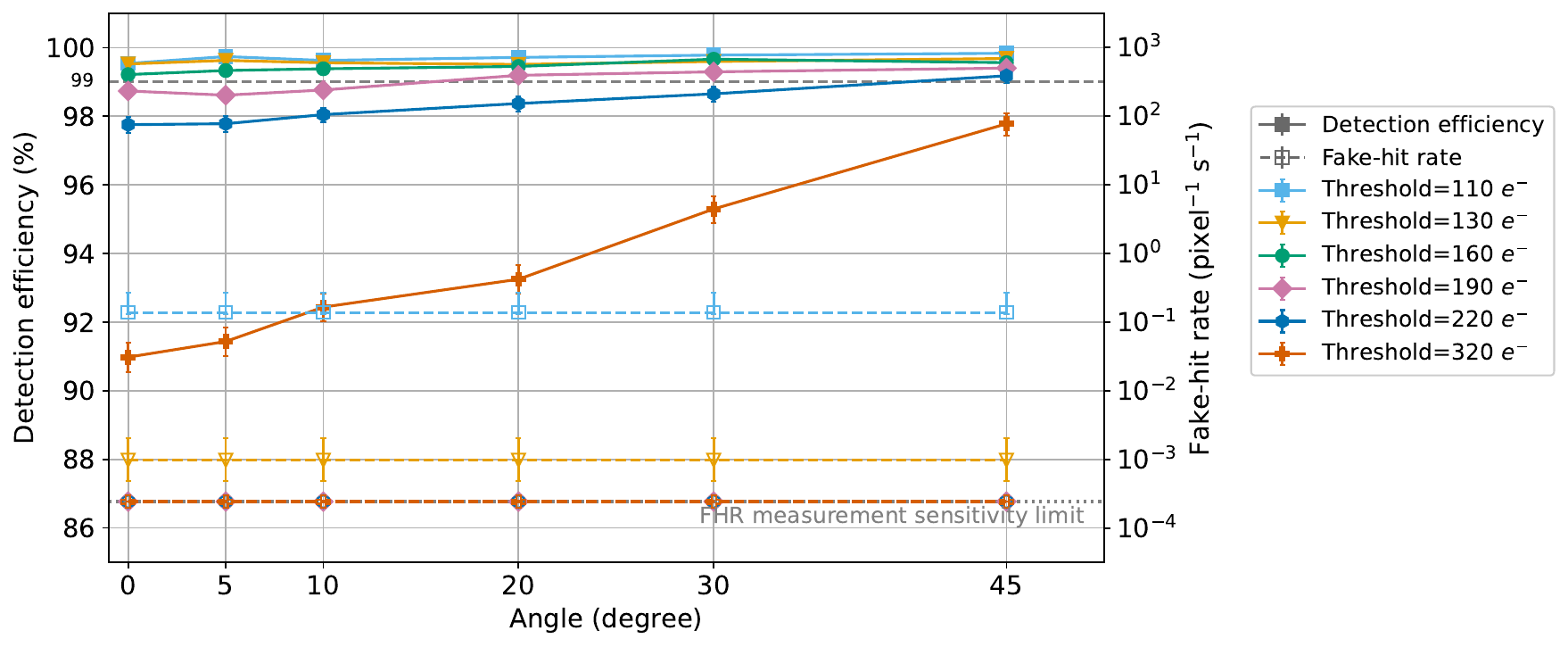}
    \caption{Detection efficiency (filled symbols, solid lines) and fake-hit rate (open symbols, dashed lines) as a function of the track incident angles for different average thresholds.}
    \label{fig:eff_vs_angle_thr}
\end{figure}

In addition to the results taken with tracks at perpendicular incidence shown in~\cite{AGLIERIRINELLA2023168589}, measurements were also taken in the angular range \qtyrange{0}{45}{\degree} at the ``nominal'' settings. Figure \ref{fig:eff_vs_thr_angle} shows the detection efficiency as a function of the threshold for several track incident angles. For all angles, the detection efficiency increases with decreasing threshold. As the track angle increases, so does the detection efficiency and the range of threshold values in which the detection efficiency is above \qty{99}{\percent}. This is further demonstrated in Fig. \ref{fig:eff_vs_angle_thr}.

\begin{figure}[!t]
    \centering
    \includegraphics[width=0.99\textwidth]    {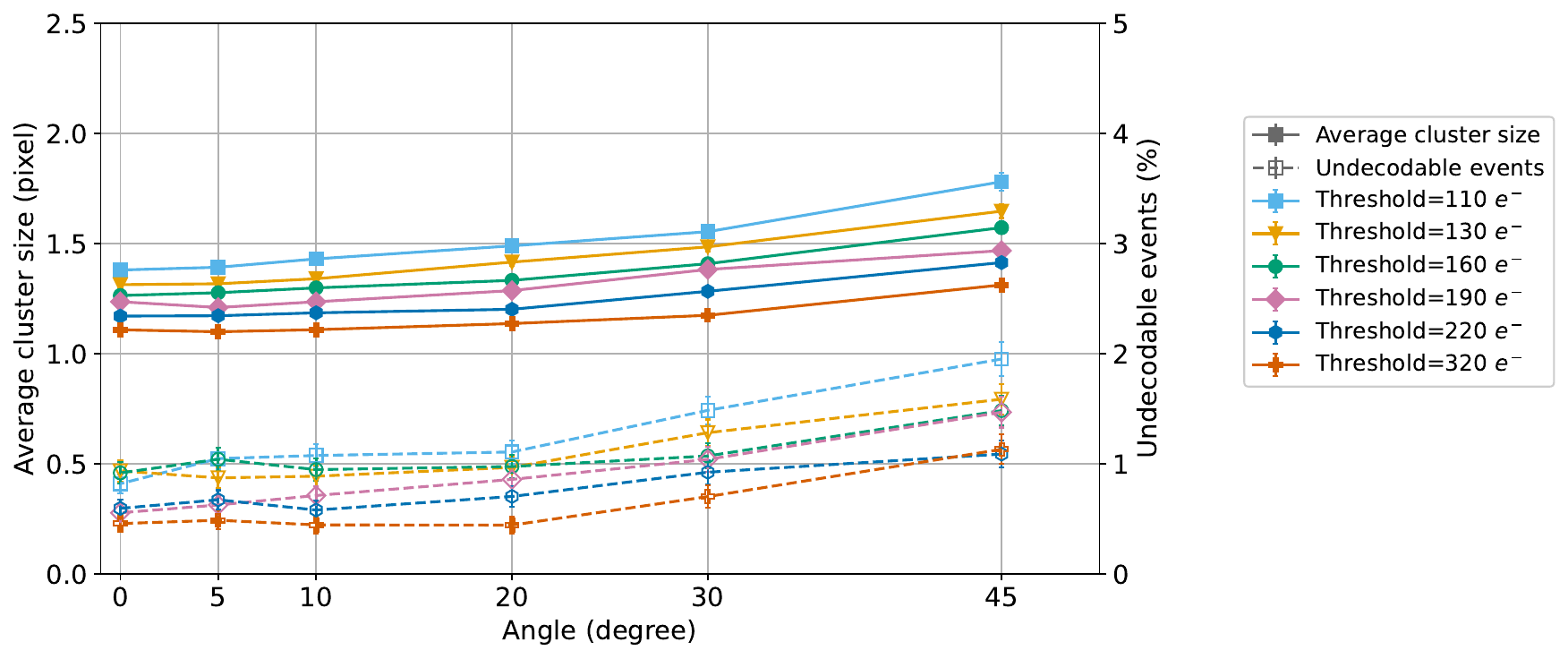}
    \caption{Average cluster size (filled symbols, solid lines) and the percentage of un-decodable hits (open symbols, dashed lines) as a function of the track incident angles for different average thresholds.}
    \label{fig:clust_size_angle}
\end{figure}

The effect of the track incident angle on the cluster size at different threshold values was also investigated and is shown in Fig. \ref{fig:clust_size_angle}. As expected, as the angle increases, so does the average cluster size, but the cluster size only changes by < 1 pixel over the whole angular range. Another important quantity to monitor is the number of undecodable events, which increases with angle, as shown in Fig. \ref{fig:clust_size_angle}. This is because, with the larger cluster size, there is a greater chance of the hits clashing on the readout line. The change in the number of undecodable events is negligible for angles up to \qty{20}{\degree}, after which there is an increase of approximately \qty{2}{\percent}.

\section{Conclusions}
Additional measurements of the Digital Pixel Test Structure MAPS prototype to those previously presented \cite{AGLIERIRINELLA2023168589} have been performed with a focus on the impact of operating the chip in low-power-consumption regimes as this is a key requirement of the ALICE ITS3 detector.

Laboratory measurements of the chip performance demonstrate that with a pixel matrix power consumption as low as \qty{15.9}{\milli\watt\per\cm^2}, there is minimal impact on the threshold RMS, noise mean and fake-hit rate compared to the ``nominal'' value of \qty{53}{\milli\watt\per\cm^2}. However, the performance degrades at lower pixel matrix power consumption values, particularly at the lowest value, \qty{5.3}{\milli\watt\per\cm^2}. There is a marginal improvement by increasing the reverse bias to $\qty{-1.8}{\volt}$. These trends are also seen in testbeam measurements where an excellent efficiency of above \qty{99}{\percent} is maintained down to \qty{10.6}{\milli\watt\per\cm^2} and a timing resolution of below \qty{100}{\ns} is achieved even for the lowest pixel matrix power consumption settings. However, the DPTS is not able to maintain an efficiency of \qty{99}{\percent} at a FHR of below \qty{0.01}{\per\pixel\per\s} while having a pixel matrix power consumption of less than \qty{15}{\milli\watt\per\cm^2}. This is expected to be improved in the final ITS3 sensor design, where, among design improvements, the pixel pitch will be larger, and thus the power density will be lower. The impact of larger pixel sizes on the performance has been studied in \cite{The:2890181,AGLIERIRINELLA2024169896} and found to be within the ITS3 requirements in terms of detection efficiency and spatial resolution.

The impact of ionising radiation was measured, showing that the operating point of the chip is affected and the performance, in terms of threshold RMS and FHR, deteriorates. These impacts on performance can be reduced through annealing, with the largest improvement occurring after 3 days.  X-ray emissions from an $^{55}$Fe source show that ionising radiation has a minimal impact on the charge collection but does impact the front-end electronics by reducing the measured ToT value of the $\text{Mn-K}_{\alpha}$ peak. In-beam measurements of the efficiency for sensors with accumulated dose up to  \qty{5}{\mega\gray} further demonstrate the negligible impact of ionising radiation on the charge collection and the reduction in the fake-hit rate with annealing time.

Further measurements investigating the sources of the FHR for the DPTS showed that the dominant contributor was the hot pixels. Applying a mask to \qty{5}{\percent} of the pixels revealed that, below a threshold of \qty{80}{\ele}, the contribution from the thermal noise dominates, and above this threshold, the contribution from random telegraph noise and undecodable events dominates. Fluorescence X-rays from manganese and tin targets were used to demonstrate the linearity of the front-end for an energy range of 1.7 to 28.5~keV. Finally, measurements of inclined tracks showed that increasing the track angle up to \qty{45}{\degree} increases the detection efficiency and average cluster size. However, the number of undecodable events also increases but remains below \qty{2}{\percent}.

Along with the previous DPTS results and the results from the Analog Pixel Test Structures \cite{AGLIERIRINELLA2024169896}, a more complete picture of the performance and capabilities of the Tower Partners Semiconductor Co.\ \SI{65}{\nm} CMOS process is now available. From this, insights into the operation and design of current and future sensors in this technology can be taken advantage of. These include the second submission in the \SI{65}{\nm} process, designated ER1, containing the first full-scale stitched sensor prototypes along with future submissions working towards the goal of a waferscale bent sensor and a fully cylindrical detector for the ALICE ITS3.

\newenvironment{acknowledgement}{\relax}{\relax}
\begin{acknowledgement}
\section*{Acknowledgements}
The measurements leading to these results have been performed at the Test Beam Facility at DESY Hamburg (Germany), a member of the Helmholtz Association (HGF). We would like to thank the coordinators at DESY for their valuable support of this test beam measurement and for the excellent test beam environment.

The authors thank the Optimato team at the Elettra Sincrotrone for providing their facility and expert support in conducting the X-ray fluorescence measurements.

ITS3 R\&D and construction is supported by several ITS3 project grants including LM2023040 of the Ministry of Education, Youth, and Sports of the Czech Republic and Suranaree University of Technology, the National Science and Technology Development Agency (JRA-CO-2563-12905-TH, P-2050706), and the NSRF via the Program Management Unit for Human Resources \& Institutional Development, Research and Innovation (PMU-B B47G670091) of Thailand.

M.~Buckland acknowledges the UKRI Science and Technology Facilities Council (STFC) (grant no. ST/W00450X/1) for supporting this work. 

P.~Becht acknowledges the support by the HighRR research training group [GRK~2058] and funding by the German Federal Ministry of Education and Research/BMBF (project number 05H21VHRD1) within the High-D consortium.

W.~Deng acknowledges the support by the National Key Research and Development Program of China (2022YFA1602103).

This work has been sponsored by the Wolfgang Gentner Programme of the German Federal Ministry of Education and Research (grant no.~13E18CHA).

The irradiation of sensors by protons was carried out at the CANAM infrastructure of the NPI CAS Rez.    
\end{acknowledgement}
%

\bibliography{references}

\end{document}